\documentclass[twocolumn,astrosymb,twocolappendix]{aastex631}

\usepackage{wasysym}
\let\iint\relax

\usepackage{amsmath}

\usepackage{xcolor}

\newcommand*{\colorboxed}{}
\def\colorboxed#1#{%
  \colorboxedAux{#1}%
}
\newcommand*{\colorboxedAux}[3]{%
  \begingroup
    \colorlet{cb@saved}{.}%
    \color#1{#2}%
    \boxed{%
      \color{cb@saved}%
      #3%
    }%
  \endgroup
}

\newcommand{\sr}{S_{\astrosun}}

\newcommand{\textapprox}{\raisebox{0.5ex}{\texttildelow}}

\shorttitle{Eclipse Solar Radius Estimation}
\shortauthors{Quaglia et al.}

\graphicspath{{./}{figures/}}

\begin{document}

\title{Estimation of the Eclipse Solar Radius by Flash Spectrum Video Analysis}

\correspondingauthor{Luca Quaglia}
\email{besselianelements@gmail.com}

\author{Luca Quaglia}
\affiliation{Sydney, New South Wales, Australia}

\author{John Irwin}
\affiliation{Guildford, England, United Kingdom}

\author{Konstantinos Emmanouilidis}
\affiliation{Thessaloniki, Greece}

\author{Alessandro Pessi}
\affiliation{Milan, Italy}

\begin{abstract}

The value of the eclipse solar radius during the 2017 August 21\textsuperscript{st} total solar eclipse was estimated to be $\sr = (959.95\pm 0.05)"$ at $1\,au$ with no significant dependence on wavelength. The measurement was obtained from the analysis of a video of the eclipse flash spectrum recorded at the southern limit of the umbral shadow path. Our analysis was conducted by extracting light curves from the flash spectrum and comparing them to simulated light curves. Simulations were performed by integrating the limb darkening function (LDF) over the exposed area of photosphere. These numerical integrations relied upon very precise computations of the relative movement of the lunar and solar limbs.

\end{abstract}


\section{Introduction}

The value of the solar radius at unit distance $\sr$ is one of the fundamental quantities needed to perform very precise eclipse computations. Some of the other required inputs are: accurate ephemerides for the position of the centres of mass of the Sun and Moon; accurate models for the orientation of the Earth and Moon; and detailed data on the topography of the Moon and Earth.

To our knowledge the solar radius is the most uncertain parameter of all these quantities. Ephemerides obtained by numerical integration, such as JPL's DE440 \citep{2021AJ....161..105P}, provide a very high level of accuracy in the positions of the Sun, Moon and Earth. For example, the residuals of the lunar distance provided by DE440 with respect to Lunar Laser Range measurements are well below $1\,m$ for the very recent past. The relative orientation of the Geocentric Celestial Reference System (GCRS) and the International Terrestrial Reference System (ITRS) is continuously measured and very accurately known (including the determination of the relationship between Universal Time and Terrestrial Time) \citep{IERS2021}. The accuracy of the orientation parameters is now below the milliarcsecond level while the accuracy of UT1 determination is well below the millisecond level. Satellite missions in the last decade have vastly improved the knowledge of the topography of the Moon \citep{2017Icar..283...70S} to better than $10\,m$ (corresponding to about $5\,mas$ at the mean geocentric distance of the Moon), allowing accurate computations of the lunar limb profile as seen by a topocentric observer.

Despite improvements in the accuracy of all other inputs needed for eclipse computations, the standard value of the solar radius at unit distance $\sr$ has remained unchanged for more than a century:
\begin{gather}
\colorboxed{red}{\sr = 959.63"}
\label{eq:auwers_radius}
\end{gather}

\noindent This is the value determined by Auwers in the late 1800's \citep{1891AN....128..361A} and it is used in all published eclipse predictions; for example in \emph{The Astronomical Almanac} \citep{Almanac2020}.

Specific domains of solar physics and astronomy may use other, more suitable, values of the solar radius based on a variety of considerations and needs. We note that, for example, in the context of conversion factors, the IAU had recommended in 2015 a nominal value for the solar radius equivalent to $\sr = 959.23"$ \citep{2016AJ....152...41P}, based on helioseismic measurements \citep{2008ApJ...675L..53H}.  

In this study we focus solely on the value of the solar radius needed for eclipse computations. That is to say, we are interested in the value of the \emph{eclipse solar radius}, the radius to be used in very precise computations that predict the internal contact times of solar eclipses to a high level of accuracy. In particular, a radius that best reproduces the visual experience of a total solar eclipse. Such a solar radius corresponds closely with the notion of \emph{complete photospheric extinction} where totality is defined as the period of time during which the photosphere is fully and unequivocally blocked by the limb of the Moon.

Auwers' value is often quoted with two decimals. This has unfortunately fostered a false sense of accuracy. The reality, highlighted by this study and by others, like \citep{2015SoPh..290.2617L}, is that the eclipse solar radius can be quite comfortably estimated to within $0.1"$ (but certainly not to within $0.01"$) and that its value is definitely larger than the standard radius by around $0.3"$ or $0.3$\textperthousand.

Among the many quantities provided by eclipse computations, the most important ones for practical applications are the internal umbral contact times for a total solar eclipse. How accurately should these times be forecast by computation? Settling on a number is highly fraught with pitfalls but just for now, as a way of providing some context, we will say accurate at the level of a tenth of a second for eclipses occurring close to the current epoch. Even if computational algorithms can be precise by accounting for many factors in the implementation of procedures and provide contact times to a tenth of a second, they can only be as accurate as the accuracy of their inputs. In this study we will only consider eclipse computations that are the most precise we can achieve nowadays so as to reduce to a minimum any other sources of inaccuracy, other than for ones depending on the uncertainty of the inputs. The latter is due mainly to how well we know the value of the solar radius as all other quantities are known to a sufficient level of accuracy.

\subsection{Selected literature}

\begin{deluxetable*}{lllll}
\tablecaption{Solar radius measurement techniques based on total solar eclipses}
\label{tbl:techniques}
\tablewidth{0pt}
\tablehead{
\colhead{Group}                   & 
\colhead{Observable}              & 
\colhead{Location}                &
\colhead{Measured Quantity}       & 
\colhead{Solar Radius derived by}
}
\startdata
Kubo et al. & 
flash spectrum & 
centreline / limits & 
light curves & 
best fitting contact times \\
IOTA & 
white light + solar filter &    
umbral path limits & 
Baily's beads times & 
best fitting Baily's beads times \\
Lamy et al. & 
white light + narrow band filter &    
centreline / limits & 
light curves & 
light curves simulation matching \\
this work & 
flash spectrum &    
umbral path limits & 
light curves & 
light curves simulation matching
\enddata
\end{deluxetable*}
\vspace{-0.85cm}

During the last fifty years several methods have been proposed and applied to measure the solar radius by taking advantage of the favourable conditions occurring during total solar eclipses. As highlighted in \citep{2015SoPh..290.2617L} the occultation happens in space and hence it is free of atmospheric effects. As measurements are performed very close to the internal contact times, when the solar disc is almost entirely occulted by the Moon, the glare of the photosphere is  minimal and the level of scattered light become neglegible. Finally, the Moon provides a reliable and long term reference, independent from any specific instrumentation. One drawback is that total solar eclipses happen infrequently, they are sometimes marred by unfavourable weather, and they only provide one chance to collect data. A summary of the main characteristics of each method are summarised in Table \ref{tbl:techniques}.

The idea of using the flash spectrum to time the contacts of total solar eclipses can be traced back to the 1950's when Kristenson applied it during the 1954 June 30\textsuperscript{th} total solar eclipse \citep{1960ArA.....2..315K} in the context of geodetic investigations to measure the size of the Earth. Kristenson collected a video of the flash spectrum and extracted light curves of photospheric light corresponding to the last Baily's beads before totality and to the first Baily's beads after totality. By looking at the inflection point of these light curves he was able to estimate the time of second and third contact.  

A more extensive use of Kristenson's technique aimed at measuring the solar radius was carried out in the 1970's and 80's by the Japanese Hydrographic Department \citep{1993PASJ...45..819K}. Once the contacts times have been obtained, the solar radius can be estimated by fitting eclipse computations to reproduce them. Kubo \emph{et al}. observed several eclipses, at times near the centreline, at times not far from the umbral shadow path limits, and estimated values for the solar radius in the range [$959.74",959.88"$], values systematically larger than the standard value of $959.63"$ used for eclipse computations.

A long running project to measure the solar radius by observing total and annular solar eclipses from the limits of the umbral shadow path \citep{1994SoPh..152...97F} is being carried out by the \emph{International Occultation Timing Association} (IOTA). Their results shows tantalising evidence of fluctuations of the solar radius over time. Positioning close to the limit has the advantage of enhancing all transient phenomena. The main technique they use is to time the disappearance and reappearance of Baily's beads by recording time-stamped videos of filtered telescopic views of the eclipse \citep{Nugent2007}. Based on eclipses observed from the ninties, IOTA estimated values for the solar radius in the range [$956.40",956.80"$] \citep{2016IAUS..320..351D}.

One of the most comprehensive recent experimental efforts to measure the solar radius is reported in \citep{2015SoPh..290.2617L}. By measuring light intensity curves collected with photometers around
the time of second and third contact and by simulating them by integrating the limb darkening function, that study has possibly obtained the most robust estimate of the solar radius $\sr$ so far: $959.99"\pm 0.06"$. The observations were conducted both at the centreline and near the umbral shadow path limits over the course of four eclipses.

Our group attempted contact timing by flash spectrum analysis during the total solar eclipses on 2009 July 22\textsuperscript{nd} (China), 2010 July 10\textsuperscript{th} (Cook Islands) and 2012 November 13\textsuperscript{th} (Australia). Several factors, mainly unfavourable weather, prevented the gathering of usable data. Finally, during the 2013 November 3\textsuperscript{rd} annular-total solar eclipse (Gabon), under almost perfect sky conditions, we succeeded in collecting a time-stamped low resolution video of the flash spectrum and we presented the results at the 2014 Solar Eclipse Conference in New Mexico, USA. On all those occasions we were stationed close to the centreline. Appreciating the advantages of collecting data from the limit of the umbral shadow path, we decided to target the southern limit for the 2017 August 21\textsuperscript{st} total solar eclipse.

\subsection{Aims}

The main objectives of this work are to present an estimate of the value of the eclipse solar radius during the 2017 August 21\textsuperscript{st} eclipse and more broadly to outline a technique that could be used at future eclipses to build a data set of eclipse solar radius measurements. This technique brings together several building blocks:

\begin{itemize}
\item eclipse flash spectrum video analysis
\item observing from the limits of the eclipse umbral shadow path
\item precise eclipse computations
\item light curves simulations
\end{itemize}

The individual building blocks have been used in the past (see Table \ref{tbl:techniques}) with the aim of measuring the eclipse solar radius but we are not aware of them being used all together. Each building block has a strong aspect and their synergy produces a measuring procedure that can be implemented with modest equipment and be undertaken by competent amateurs, at least where it concerns the first two building blocks.

In the Methodology section we will extensively discuss each of those building blocks and how they come together into a coherent procedure. In the Results section we will first present an account of the naked eye view of the eclipse as seen from the limit of the eclipse shadow path. This description will help with a general understanding of the evolution of the flash spectrum. Then we will provide an initial estimate of the eclipse solar radius by estimating the duration of totality based only on visual inspection of the flash spectrum video. This will provide clear evidence that Auwers' radius (Equation \ref{eq:auwers_radius}) is notably too small to reproduce the observations. Finally, we will estimate the eclipse solar radius with a more quantitative procedure based on extracting light curves from the flash spectrum video and matching simulated light curves with observed curves.      

\section{Methodology}

Estimation of the value of the eclipse solar radius is performed by analysing a video of the flash spectrum recorded at the very limit of the umbral shadow path. The method hinges on extracting light curves from the last and first Baily's beads around the times of second and third contacts and on comparing them to simulated light curves. 

The temporal evolution of the light curves is closely dependent on the duration of totality which is, in turn, very sensitive to the value of the solar radius $\sr$. Matching simulated light curves to observed ones is more stringent than just estimating the duration of totality or the internal contact times. This is because the complete evolution of the light curves needs to be reproduced, not just at two measurement points. 

\subsection{Recording the flash spectrum video}

The video\footnote{\url{https://youtu.be/eTqblNG-BHw}} of the flash spectrum was recorded\footnote{Original video recording is available on request.} using a Canon EOS 6D Digital camera with a Canon EF 70-200mm f/4L USM lens. The focus and the exposure were both set to manual and we chose an exposure time of $1/60\,s$ at an ISO setting of $100$. A transmission diffraction grating, etched with $235\, lines/mm$, was mounted in front of the lens to separate the eclipse light into its spectral components.

The camera produces compressed video using the AVC/H.264 codec operating on full-range [0,1] image data in the $Y'UV$ colour space with 4:2:0 chroma subsampling. The resulting video stream is stored within a MOV media container. The video has a frame size of $1920\times1080$ pixels and a frame rate of $23.976\,fps$.

DSLR cameras are not specifically designed for extracting light curves but they are relatively inexpensive, readily available and they have been used successfully for astronomical photometric work \citep{AAVSO2016}. An important aspect to highlight is that the internal CMOS sensor is linear for intensities away from saturation and above dark current noise \citep{2016AdSpR..57..509P}. These are indeed the characteristics of the video from several tens of seconds before totality to a couple of tens of seconds after totality, which is the period of interest for the video analysis.

However, one other aspect that needs to be accounted for is the presence of a non-linear gamma adjustment in the colour space of the images stored in the video stream. This adjustment originates in the camera's processing of the image data through the $sRGB$ colour space prior to their conversion to $Y'UV$ \citep{Canon6D}. The transformations required to undo this adjustment and obtain linearised luminance information are described in Section \ref{sec:gamma}.

Watching this video of the flash spectrum is an integral part of reading this article. We invite the reader to watch the video several times -- it is only by going back and forth between the details presented in this article and the video that our analysis will fully take shape.

\subsection{Observing from the limit}

The video of the flash spectrum was recorded during the 2017 August 21\textsuperscript{st} total solar eclipse from a location just south of Vale, Oregon, USA with WGS84 geodetic coordinates: 

\begin{equation}
\setlength{\arraycolsep}{1pt}
\begin{array}{r c r r r l}
\lambda &=& 117^{\circ} & 13' & 09.8" & W\\
\varphi &=&  43^{\circ} & 57' & 10.9" & N\\
      h &=& 711\, m
\label{eq:coordinates}
\end{array}
\end{equation}

\begin{figure*}[tb!]
\plotone{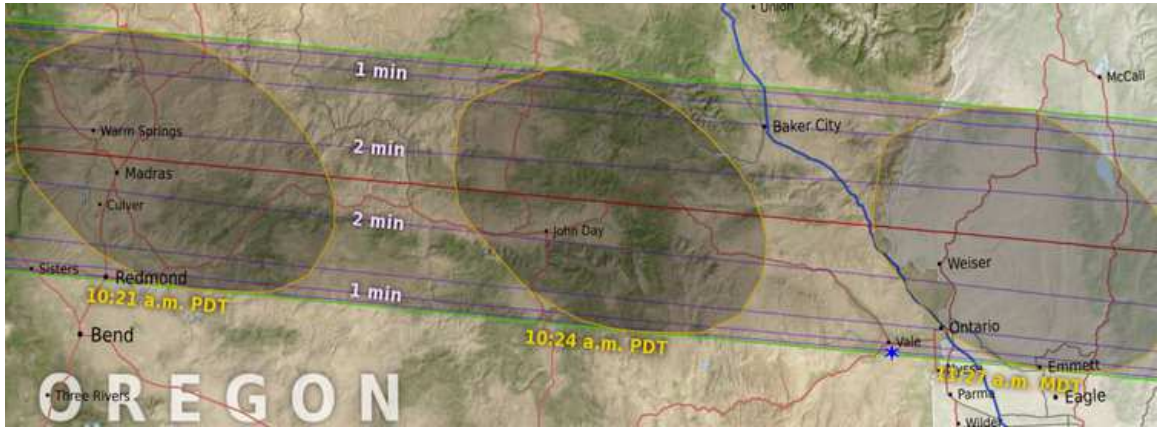}
\caption{The 2017 August 21\textsuperscript{st} total solar eclipse was observed from a location (indicated by a star) south of Vale, just a few hundred metres inside the southern limit of the eclipse (original image by Ernie Wright, Scientific Visualization Studio, NASA, \url{https://svs.gsfc.nasa.gov/4552}).}  
\label{fig:location}
\end{figure*}

\noindent This location was very close to the southern limit of the umbral shadow path of the eclipse as Figure \ref{fig:location} shows.

\begin{figure*}[tb!]
\gridline{
        \fig{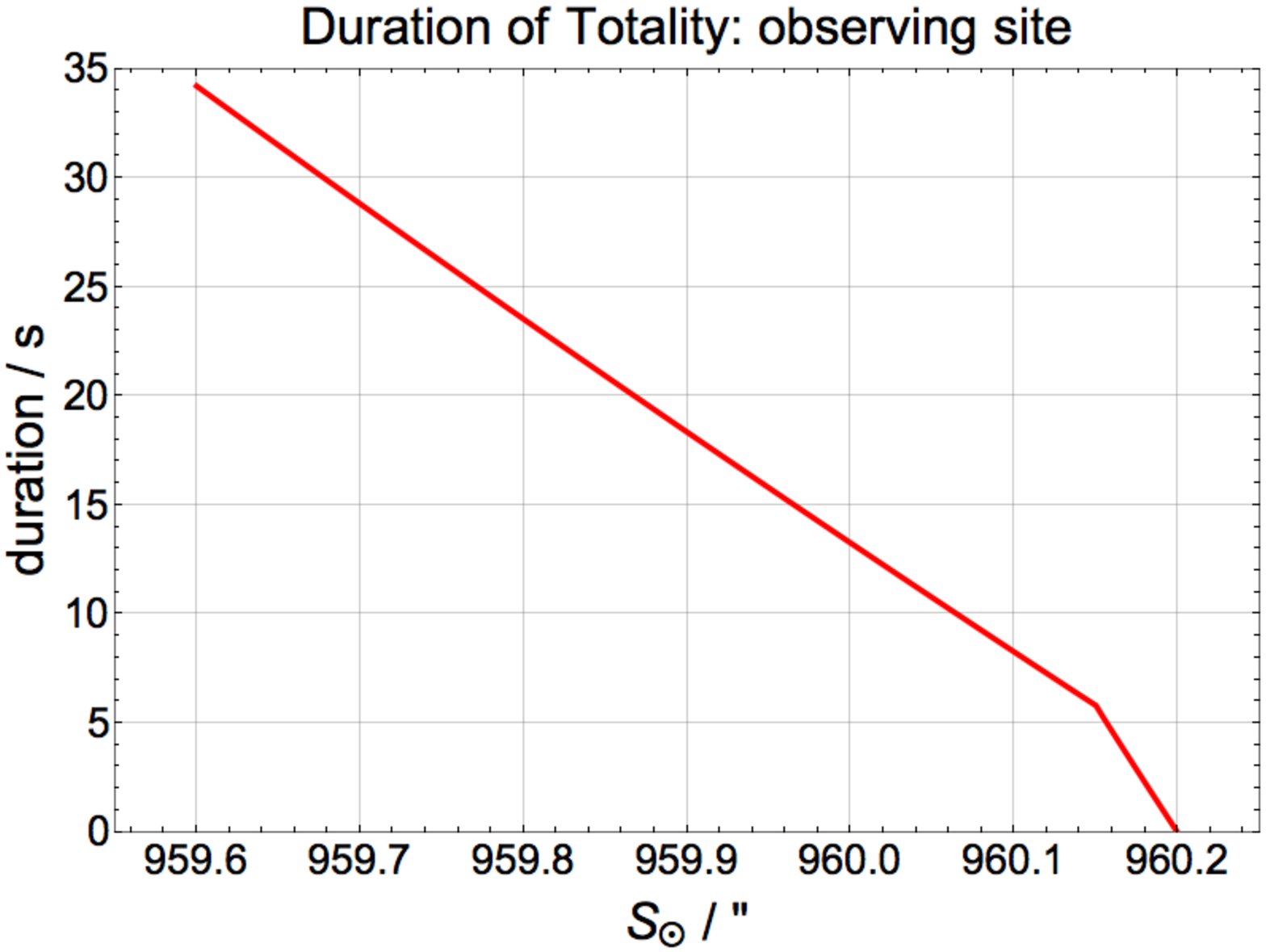}{0.32\textwidth}{(a)}
        \fig{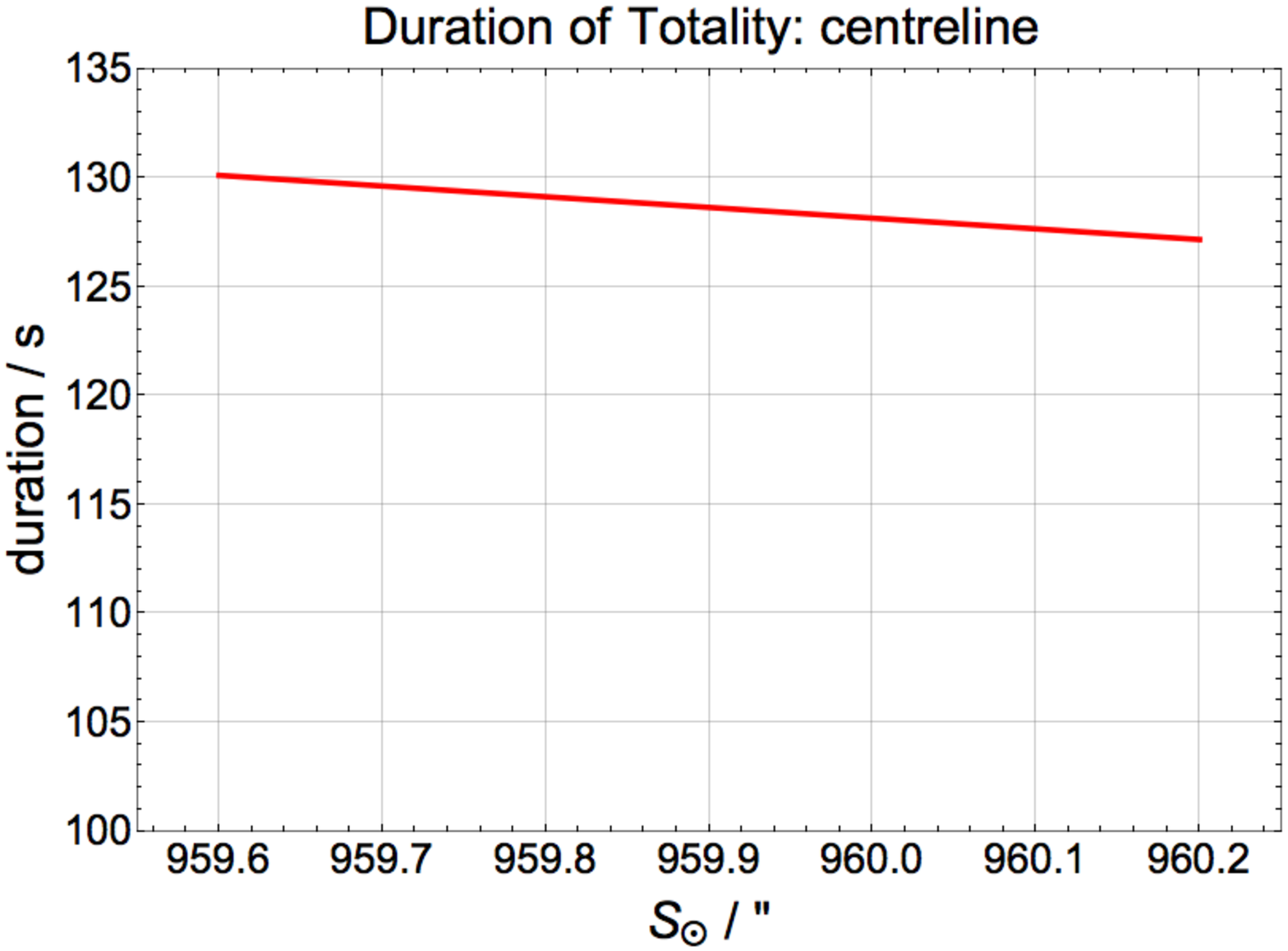}{0.32\textwidth}{(b)}
        \fig{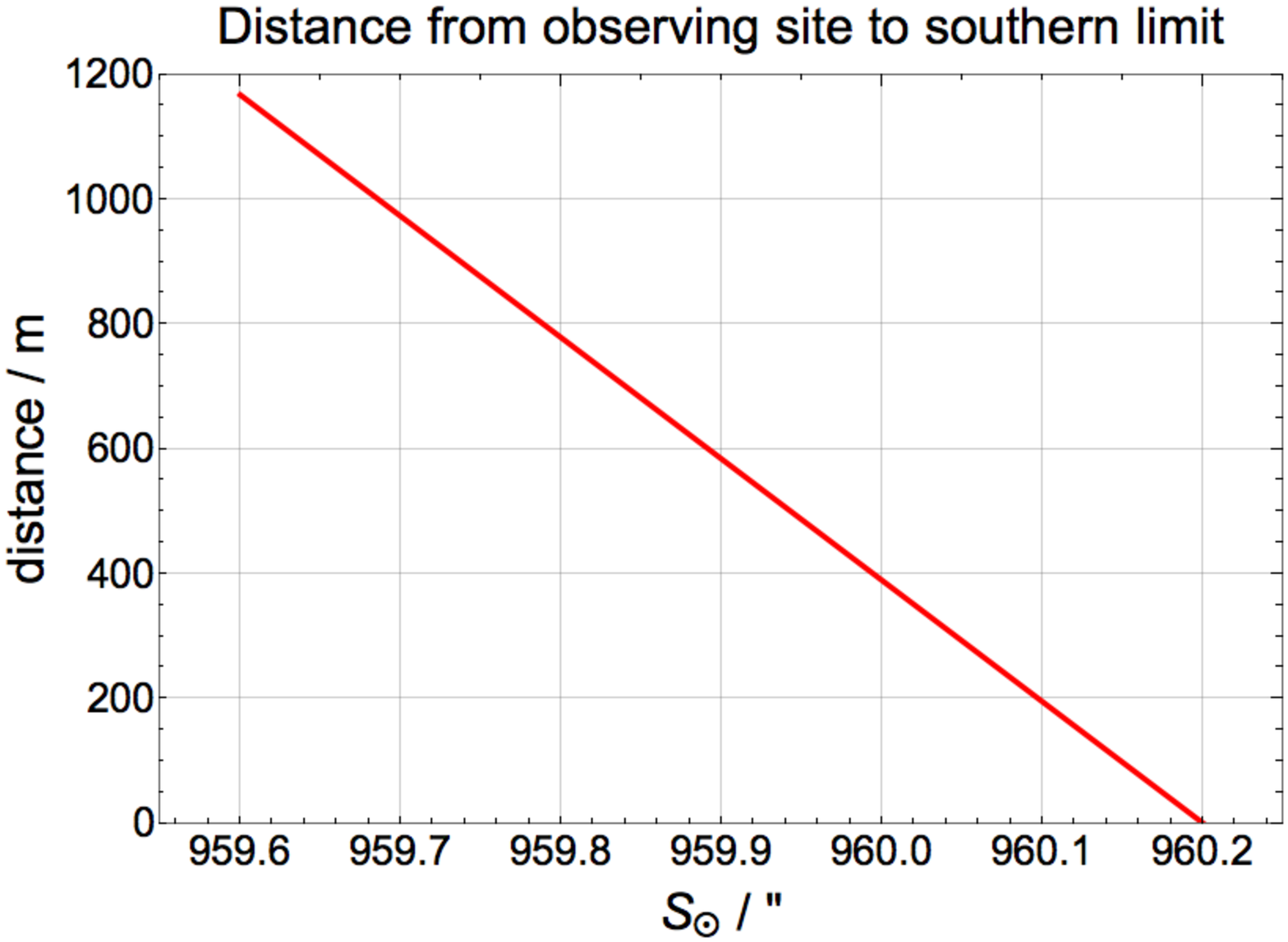}{0.32\textwidth}{(c)}
         }
\caption{Dependency on the value of the solar radius $\sr$ of some eclipse related quantities: (a) duration of totality at the observing site; (b) duration of totality at a location on the centreline with mid-eclipse happening at the same time as the observing site; (c) distance between the observing site and the southern limit of the umbral shadow path. (J. Irwin)
}
\label{fig:duration_distance}
\end{figure*}

Positioning very close to the edge of the umbral shadow path enhances all transient phenomena and greatly increases the sensitivity of the duration of totality to the value of the solar radius. Figure \ref{fig:duration_distance} (a,b) shows the predicted duration of totality as a function of the solar radius at the observing site and at a site on the centreline experiencing mid-totality at the same time. We can see that, very close to the umbral shadow path limits, the duration of totality is noticeably more sensitive to the adopted value of the solar radius $\sr$ than at the centreline. If we compare the durations computed assuming the standard value $\sr = 959.63"$ and an increased value $\sr = 960.00"$, similar to the one found in \citep{2015SoPh..290.2617L}, on the centreline the drop in duration is only $1.8\,s$ while at the observing site the drop is $19.3\,s$. We note that the bend in the plot in Figure \ref{fig:duration_distance} (a) is not a discretisation error (data are computed every 0.01") but an actual feature due to the shape of the lunar limb profile.

Figure \ref{fig:duration_distance} (c) also shows the distance between the observing site and the umbral shadow path limit as a function of the solar radius. If $\sr = 959.63"$, this distance is \textapprox$1200\,m$, while with an increased value $\sr = 960.00"$, it drops to less than $400\,m$, a substantial change.

\subsection{The eclipse flash spectrum}

The spectrum of solar light is continuous under everyday circumstances. The photons originating from thermonuclear reactions inside the Sun slowly make their way to the surface and radiate away at all wavelengths (being visible to the naked eye between about $400\,nm$ and $700\,nm$). While travelling through the solar atmosphere photons at specific wavelengths are absorbed by specific elements like Hydrogen, Magnesium, Sodium and Iron. These absorption phenomena create dark lines in the continuous spectrum and are the hallmarks of the presence of those elements in the solar atmosphere.

\begin{figure*}[tb!]
\plotone{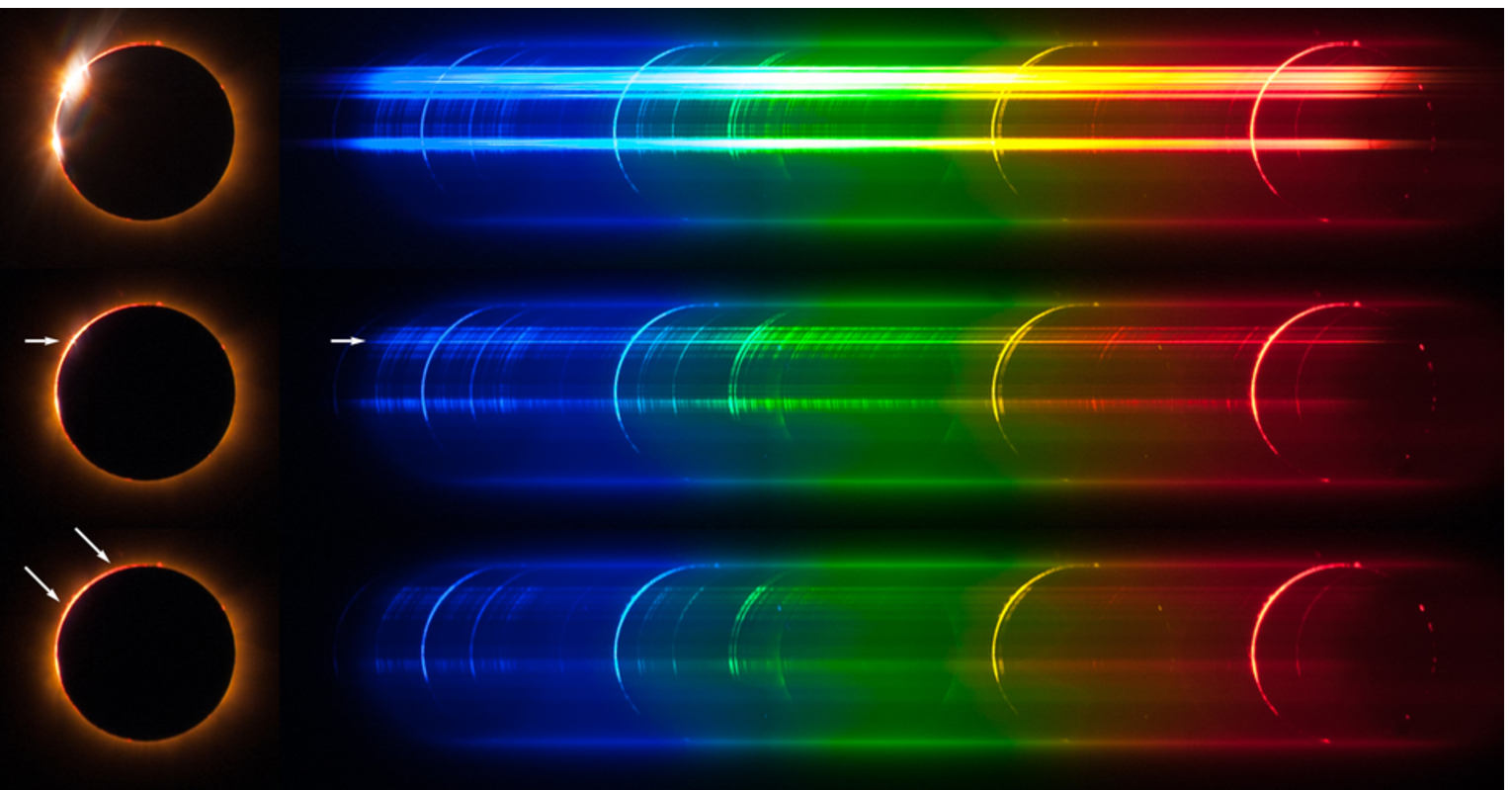}
\caption{The eclipse and its flash spectrum captured in a sequence of still images at second contact during the 2013 November 3\textsuperscript{rd} annular-total eclipse. Both the white light view of the eclipse and its spectrum were captured on the \emph{same} still image, so the resolutions are identical. As the eclipse becomes total, the continuous photospheric spectrum vanishes and the emission spectrum from the solar atmosphere comes fully into view. The last Baily's bead and its corresponding continuum are indicated by arrows in the middle frame. In the last frame we highlight with arrows that some parts of the chromosphere shine with almost white light (not to be confused with photospheric light) and some have a deep purple-red colour. (K. Emmanouilidis)}  
\label{fig:flashspectrum_sequence}
\end{figure*}

The spectrum of solar light changes rather dramatically just a handful of seconds before second contact of a total solar eclipse and reverts back to its normal state just a handful of seconds after third contact. Figure \ref{fig:flashspectrum_sequence} displays an example of such a transition during the 2013 November 3\textsuperscript{rd} annular-total eclipse. The usual intense continuous spectrum gives way, for some brief precious moments, to the mesmerising flash spectrum. When the Moon has almost entirely covered the photosphere, the intensity of the solar continuum plummets rapidly before vanishing and the much fainter light emitted by elements in the chromosphere and the inner corona comes into view. The continuous spectrum with dark absorption lines is replaced by a discrete emission spectrum with a few intense arcs and a myriad of fainter ones.

The emission arcs emitted by the chromosphere are at their brightest just after second contact. As totality progresses and the Moon covers the chromosphere (at least as seen from around the centreline), these arcs fade away and only the faint and ghostly diffuse spectrum of the corona remains visible. The emission arcs come back towards the end of totality and they intensify to become their brightest again just before third contact. Soon afterwards the photospheric spectrum reappears, very rapidly brightens up and hides the flash spectrum from view until the next total solar eclipse occurs. Traditionally, the name flash spectrum has been reserved for the highly dynamic spectrum visible for just a few seconds near second and third contact when the Moon has fully covered the photosphere but has not yet occulted the chromosphere. It should be noted though that during intrinsically brief total eclipses (with a magnitude $A$ very close to $1$) the flash spectrum might remain in view for most, if not the entire, duration of totality (this was the case, for example, for the 2013 November 3\textsuperscript{rd} annular-total eclipse).

\begin{figure*}[tb!]
\gridline{\fig{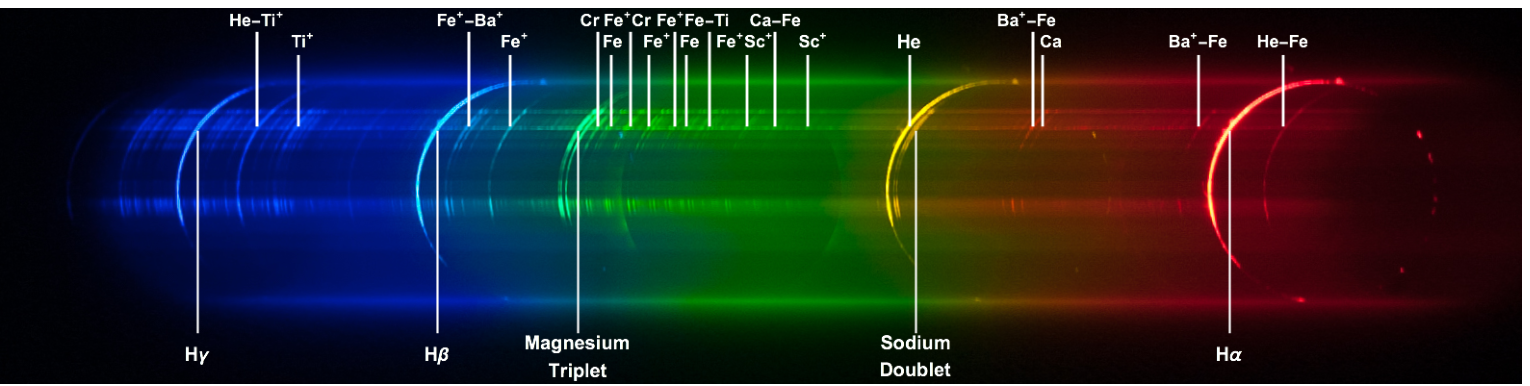}{0.85\textwidth}{(a)}}
\gridline{\fig{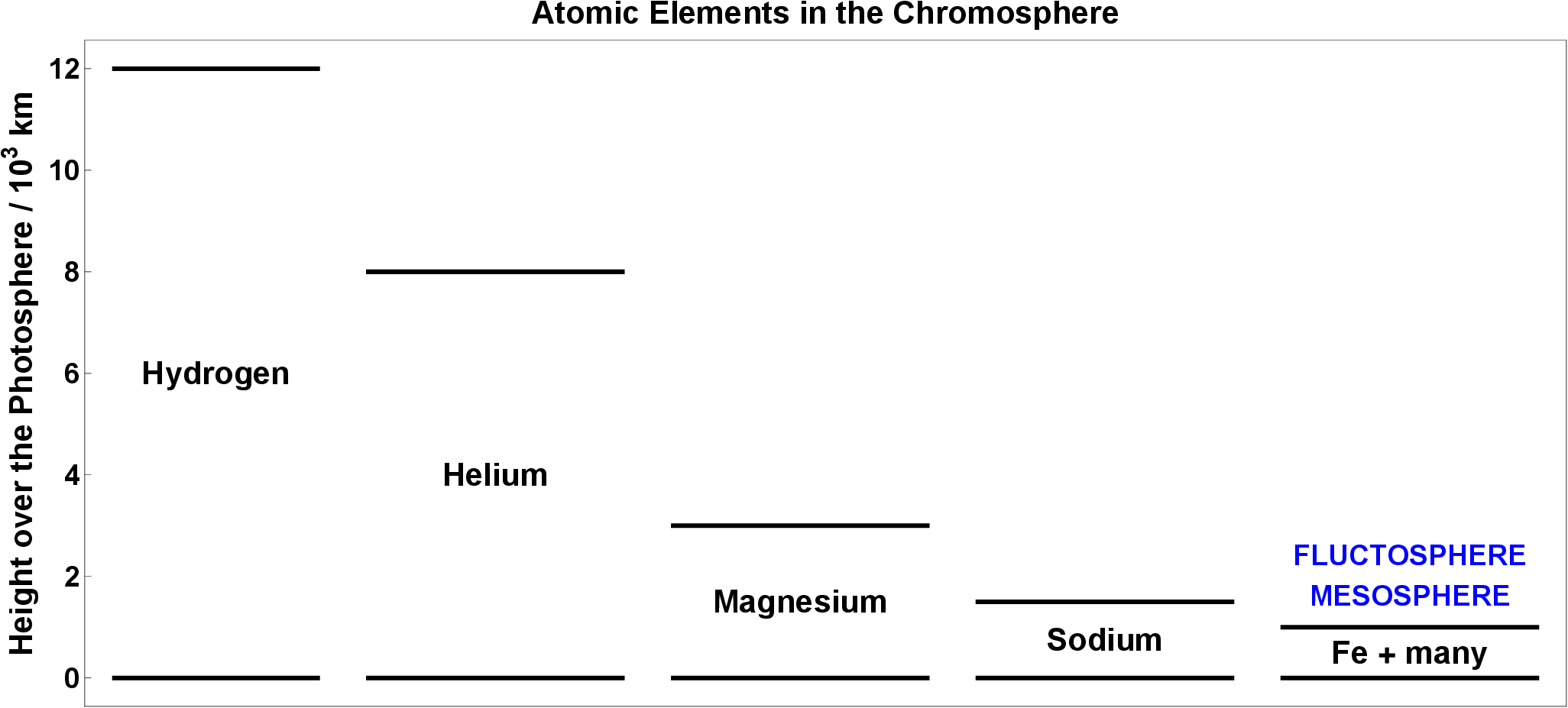}{0.5\textwidth}{(b)}}
\gridline{\fig{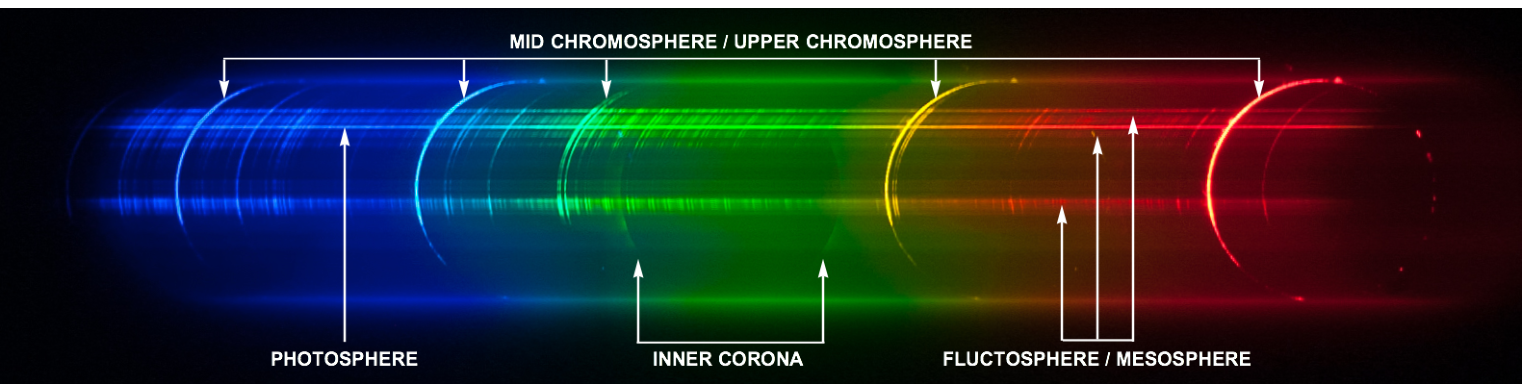}{0.85\textwidth}{(c)}}
\caption{A detailed view of the flash spectrum. (a) The main emission lines are labelled with the elements that produce them. (b) Simplified view of the distribution of elements in the chromosphere (inspired by \citep{1947ApJ...105....1M}). All elements are present, in measurable quantities, from the surface of the photosphere up to a certain height: Hydrogen extends up to beyond $10^{4}\,km$, while Iron and many other elements are present in detectable amounts only very close to the surface. (c) The photosphere generates an intense continuous spectrum (here just a thin line due to the last Baily's bead). The inner corona generates a ghostly diffuse spectrum. The mid and upper chromosphere only emits a few intense lines. The lower part of the chromosphere also emits a forest of faint lines besides the main intense ones, here predominately concentrated in the two lunar valleys of the last pair of Baily's beads. (K. Emmanouilidis)}
\label{fig:flashspectrum_details}
\end{figure*}

Figure \ref{fig:flashspectrum_details} (a) shows the origin of the main emission lines in the flash spectrum. Electronic transitions of various elements, in the neutral state or in ion form, emit photons at specific wavelengths. Prominent are the emission arcs of the Balmer series of Hydrogen ($H\text{-}\alpha$, $H\text{-}\beta$, $H\text{-}\gamma$...), some emission arcs from Helium, the Magnesium Triplet and the Sodium Doublet. The forest of faint arcs is due to a host of elements like Iron, Calcium, Barium, Titanium, and Chromium. Superposed on the discrete spectrum is the very faint, almost ghostly, circular-shaped diffuse spectrum of the inner corona.

To provide a deeper explanation of the structure of the flash spectrum we need to make some assumptions on the distribution of elements in the chromosphere. We wish to express the caveat that we are not solar physicists and we do not claim to be depicting the most detailed available knowledge. We just want to give an approximate idea of what could be going on in the solar chromosphere to have more tools to understand the flash spectrum. Based on investigations of the chromosphere by S.A.Mitchell \citep{1947ApJ...105....1M} we imagine the distribution of elements as in Figure \ref{fig:flashspectrum_details} (b). We start by pointing out that the location where the photosphere ends and the chromosphere begins is not well defined as it is not precisely known where the top of the chromosphere lies. The Sun is not a solid body and the top of the chromosphere is turbulent and the place of occasional large prominences. Nevertheless, for simplicity's sake we provide some approximate indications of where the light emitting elements are located. Hydrogen may be present, in probably abundant density, up to a height of \textapprox$12000\,km$ above the surface of the photosphere, Helium up to \textapprox$8000\,km$, Magnesium up to \textapprox$3000\,km$, Sodium up to \textapprox$1500\,km$, and Iron and a host of other elements up to a height of \textapprox$1000\,km$. Given these heights, we can infer that, at second contact for example, a wider arc of light will be emitted by Hydrogen, extending outwards by at least $12000\,km$, than by Iron which extends outwards by only $1000\,km$. We do not make any hypotheses on the density distribution of specific elements within their respective layers; it is probably inhomogeneous.

From geometrical reasoning we expect to see long arcs at the emission wavelengths of Hydrogen: $656.3\,nm$  ($H\text{-}\alpha$, red), $486.1\,nm$ ($H\text{-}\beta$, aqua), $434.0\,nm$ ($H\text{-}\gamma$, blue). Almost as long is the arc at the wavelength of Helium: $587.6\,nm$  (yellow). We expect shorter arcs in correspondence to the Magnesium Triplet: $516.7$, $517.3$, $518.4\,nm$ (green) and the Sodium Doublet: $589.0$, $589.6\,nm$  (yellow). Iron and other elements emit a forest of much shorter arcs at a wide range of wavelengths. We need to note that, due the nature of the electronic transitions involved, not all emission lines have the same strength -- some are very intense, like the $H\text{-}\alpha$ and the Helium line at $587.6\,nm$.  

Armed with this knowledge we can try to explain the naked eye appearance of the chromosphere (refer to the still image at the bottom of Figure \ref{fig:flashspectrum_sequence}). As most of the chromosphere above \textapprox$3000\,km$ is dominated by the emission of Hydrogen and Helium, when we combine the respective red, yellow, aqua and blue emission lines, we get that deep purple-red colour that is so characteristic of many images of the chromosphere. However, much closer to the photosphere, besides the emissions of Hydrogen, Helium, Magnesium and Sodium, we have a myriad of emission lines by Iron and other elements. These lines are discrete but they are very densely distributed across most of the visible spectrum. The combined effect of all these emission lines is to generate a whitish light very close in aspect to the light emitted by the photosphere. This light is not to be confused with the remnants of the photosphere. 

That part of the chromosphere closest to the photosphere has a whitish colour that is distinctively different from the pinkish-purple one of the majority of the chromosphere lying higher up. In the bottom frame of Figure \ref{fig:flashspectrum_sequence} we have pointed out these subtle differences in the visible chromosphere -- some areas are intense and almost white and some are less intense and more purple-red in colour. The corresponding spectrum clearly explains the origin of such differences in colour. Even if it does not seem to be in common use, we have come across the terms \emph{fluctosphere} \citep{2009SSRv..144..317W} and \emph{mesosphere} \citep{2012IJMPS..12..405S} to describe this lowest-lying layer that is the source of the myriad of faint emission lines. Because the fluctosphere/mesosphere emits a quasi-continuum of white light, albeit with fainter intensity than the photosphere, it is troublesome to disentangle the location of the very edge of the photosphere from the mesosphere in white light images.

The flash spectrum comes to the rescue as it allows, at least in principle, separation of the light coming from the photosphere, the fluctosphere, the rest of the chromosphere, and from the corona, as indicated in Figure \ref{fig:flashspectrum_details} (c). In practise, this statement might seem quite over-optimistic. But it can at least be said that, for the same imaging resolution, the flash spectrum provides a far greater level of detail than the corresponding white light image (see Figure \ref{fig:flashspectrum_sequence}).

\pagebreak

\subsection{Precise eclipse computations}

The eclipse computational model relies on the latest ephemerides and accounts in a very precise way for all the complexity of the lunar topography and of the orientation of the celestial and terrestrial reference systems via robust algorithms and procedures.

Modelling of the lunar topography is based on the gridded elevation data sets SLDEM-256 and LDEM-128 originating from  measurements by the \emph{Lunar Orbiter Laser Altimeter} (LOLA) on board the \emph{Lunar Reconnaissance Orbiter} \citep{NASA2021}. The former data set is a special reduction of the LOLA observations in conjunction with photogrammetric data obtained from the \emph{Kaguya} mission which resulted in a more accurate and reliable digital elevation model \citep{2016Icar..273..346B}. But it is available only for latitudes within $60^{\circ}$ of the lunar equator, therefore we have used the latter data set to cover latitudes outside this range. These data are referenced to the mean-earth/mean-axis (ME) lunar orientation frame as defined for JPL's DE421 planetary and lunar ephemeris \citep{JPL2021}. Additionally, the data reference the centre of mass of the Moon so that they can be used directly with the lunar ephemeris.

However, we have used the later DE430 ephemeris to provide better accuracy for the positions of the Sun, Moon, and Earth. We also used the lunar orientation data contained in this ephemeris but note that there is effectively only a few metres difference between the ME orientations provided by DE421 and DE430 \citep{2018CeMDA.130...22A}, which is negligible for our purposes. We also note that updated ephemerides released by JPL since DE430 do not provide any data for the ME frame, only for the principal axis (PA) frame. The difference between the PA and ME frames is typically around $850\,m$ on the surface of the Moon and cannot be neglected in computing reliable lunar limb profiles.

For Earth orientation we used the "IAU-2006" model, comprising IAU-2006 precession, IAU-2006A adjusted nutation, and IAU-2006/IAU-2000A sidereal time \citep{IERS2010}. Combined with measured Earth orientation parameters for UT1-UTC, polar motion, and the celestial pole offsets \citep{IERS2021}, this model provides the orientation of the ITRS relative to the GCRS to better than $1\,mas$ ($<0.1\,m$ on the ground). Therefore, given accurate coordinates for the observation site, this modelling ensures accurate topocentric positioning of the Sun and Moon.

The prediction of apparent topocentric positions follows a fully relativistic procedure involving adjustments for (a) light time to the body in question; (b) gravitational bending of the light path by the mass of the Sun; and (c) planetary aberration due to the barycentric velocity of the observer. The latter is used in preference to stellar aberration as it produces, in particular, a more accurate prediction for the distance of the Moon and therefore a better estimate of the apparent lunar semidiameter \citep{Irwin2019}.

Our computational model does not depend, like the traditional method, on Besselian elements \citep{Supplement2012} to compute nominal predictions and then on applying limb corrections \citep{1983JBAA...93..241H} to account for the Moon's topography. Instead, it performs its computations directly with the solar limb and the lunar limb profile to find times of contact and other related quantities. No approximating adjustments are involved with these computations.

The way the solar and lunar limbs move with respect to each other determines how the flash spectrum evolves around the times of second and third contact. To assess these dynamics for the 2017 August 21\textsuperscript{st} total solar eclipse as seen from our observing site we have computed accurate time-dependent polynomials providing the apparent topocentric semidiameters of the Sun and Moon and the relative position of their centres of mass. The topocentric lunar limb profile is also computed. All these data can be derived from the eclipse computational model. 

The polynomials are centred at the time:

\begin{equation}
T_{0} = 17^{h}\, 25^{m}\, 55.1^{s}\, \text{UTC}
\label{eq:reference_time}
\end{equation}

\noindent which corresponds to the time of mid-eclipse at the coordinates given in Equation \ref{eq:coordinates} and $\sr = 960.00"$. All polynomials are of the form:

\begin{equation}
c_{0} + c_{1}\,t + c_{2}\,t^{2} + c_{3}\, t^{3}
\end{equation}

\noindent where the coefficients $c_{i}$ are given in Table \ref{tbl:polynomials} and $t$ is the time in minutes from the reference time $T_{0}$. The polynomial for the apparent topocentric solar semidiameter corresponds to $\sr = 960.00"$. For any other solar radius within \textapprox$10"$ of this value the polynomial can be scaled by the dimensionless factor:

\begin{equation}
\left(\frac{\sr}{960"}\right)   
\end{equation}

\noindent to give the corresponding semidiameter to better than $0.1\,mas$. The apparent lunar semidiameter corresponds to a lunar radius of $1738.091\,km$. The right-handed reference frame for the $(x,y)$ coordinates is the same as for the lunar limb, with the $y$-axis pointing in the direction of the lunar north pole.

\begin{deluxetable*}{lDDDD}
\tablecaption{Coefficients of the polynomials}
\tablehead{
\colhead{Quantity} & 
\multicolumn2c{$t^{0}$}  & 
\multicolumn2c{$t^{1}$}  & 
\multicolumn2c{$t^{2}$}  &
\multicolumn2c{$t^{3}$}  \\
\colhead{}                    & 
\multicolumn2c{$"$}           & 
\multicolumn2c{$" / min$}     & 
\multicolumn2c{$" / min^{2}$} &
\multicolumn2c{$" / min^{3}$}
}
\decimals
\startdata
Topocentric solar semidiameter & $949.068186$ & $+0.000211$ & $-2.090\times10^{\text{-}7}$ & $-3.862\times10^{\text{-}10}$ \\
Topocentric lunar semidiameter & $975.800143$ & $+0.022788$ & $-8.569\times10^{\text{-}6}$ & $-9.594\times10^{\text{-}8}$  \\
X-coordinate of Sun's centre of mass &  $-0.042862$ & $+24.114466$ & $-1.1441\times10^{\text{-}2}$ & $+2.6\times10^{\text{-}5}$  \\
Y-coordinate of Sun's centre of mass & $-25.299671$ & $-2.071611$ & $+0.9225\times10^{\text{-}2}$ & $-0.6\times10^{\text{-}5}$ 
\enddata
\label{tbl:polynomials}
\tablecomments{Solar semidiameter for $\sr = 960.00"$. Lunar semidiameter for a radius of $1738.091\,km$. The coordinates of the Sun are relative to the centre of the Moon. The polynomials are valid over the interval $t = \pm 1\,min$. (J. Irwin)}
\end{deluxetable*}
\vspace{-0.85cm}

\begin{figure}[tb!]
\plotone{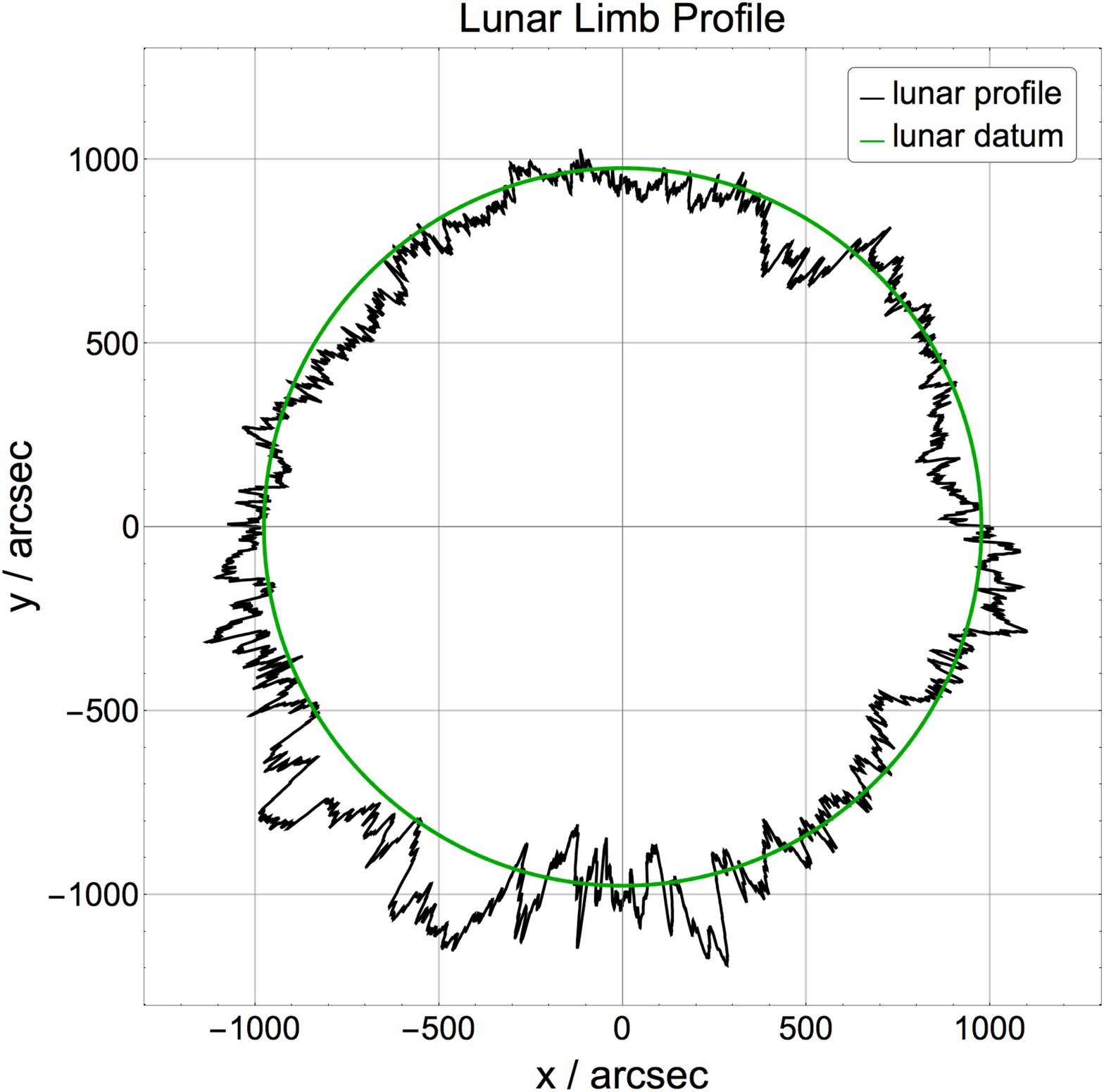}
\caption{Lunar limb profile as seen from south of Vale during the 2017 August 21\textsuperscript{st} total solar eclipse. The departure of the limb from the lunar datum is magnified $100$ times, highlighting the details of mountains and valleys. In this view, lunar north is up. For topocentric libration: $L = +5.200^{\circ}$, $B = -0.172^{\circ}$, $R = 367399.181\,km$ and a profile resolution of $0.02^{\circ}$. (J. Irwin)}
\label{fig:lunar_limb}
\end{figure}

The lunar limb profile for an observer at the coordinates given at Equation \ref{eq:coordinates} is presented in Figure \ref{fig:lunar_limb}. The mountains and valleys on the lunar limb, despite having a height of just $2-3"$ (compared with the radius of the lunar datum of around $976"$), have a great impact on determining the times of second and third contact. The contact angles of second and third contact are always at the bottom of a lunar valley. These two valleys, together with nearby valleys, are the cause of the stunning Baily's beads seen on the lead up to second contact and just after third contact. In correspondence with these Baily's beads we can see bands of photospheric continuum in the flash spectrum.

\begin{figure*}[tb!]
\plotone{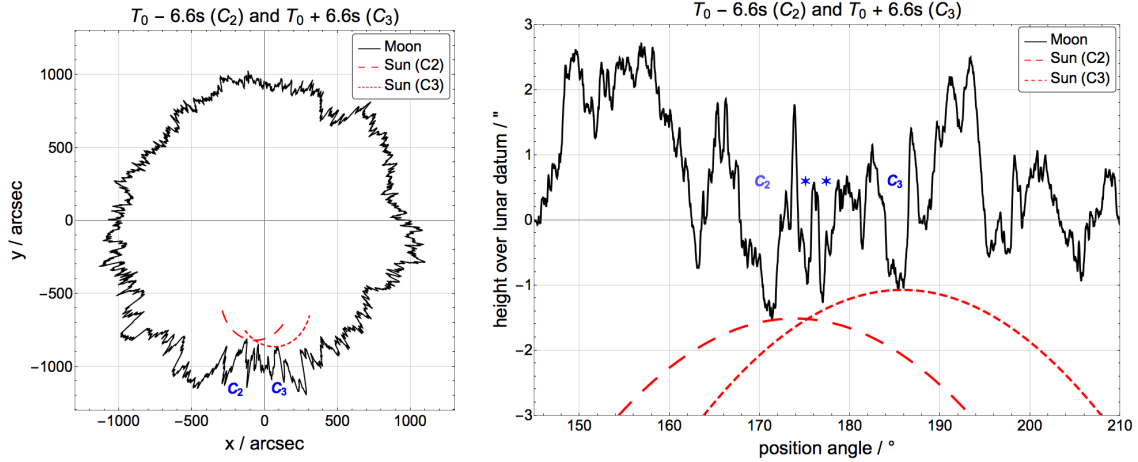} 
\caption{Diagrams displaying the relative position of the lunar and solar limbs at second and third contact. The assumed solar radius $\sr$ is $960"$. Besides indicating the valleys where second (C2) and third (C3) contact occur, the intermediate double valley is also highlighted with stars. (J. Irwin)}
\label{fig:dynamic_C2C3} 
\end{figure*}

The position of the solar limb with respect to the lunar limb at second and third contact is presented in Figure \ref{fig:dynamic_C2C3}. Totality started and ended at the bottom of two valleys in the lunar southern polar region. These two valleys are very close and are separated by a narrow and less deep double valley (indicated by stars in the diagrams). The solar limb passed just below the bottom of this double valley during totality. These three valleys generate some interesting features in the flash spectrum. We expect the photospheric continuum to disappear in the valley labelled $C_{2}$ and to reappear in the valley labelled $C_{3}$. Photospheric continuum does not shine in the intermediate double valley but the presence of this double valley is seen in the flash spectrum, as we will show later on.

\begin{figure*}[tb!]
\plotone{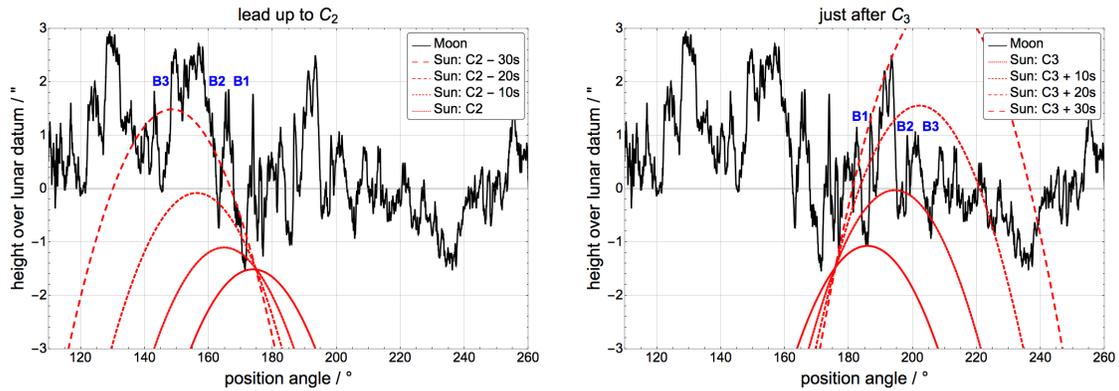}
\caption{Diagrams displaying the relative position of the lunar and solar limbs on the lead up to second contact and just after third contact. The assumed solar radius $\sr$ is $960"$. The location of valleys generating Baily's beads is indicated. For clarity, only the lunar limb near the southern pole region and the solar limb are displayed. (J. Irwin)}
\label{fig:dynamic_before_C2_after_C3}
\end{figure*}

We can also look at the mutual movement of the solar and lunar limbs on the lead up to second contact and just after third contact. These dynamics are presented in Figure \ref{fig:dynamic_before_C2_after_C3}. In this figure we have indicated (with the tags $B1$, $B2$ and $B3$) the valleys generating the last three Baily's beads before $C_{2}$ and the first three Baily's beads after $C_{3}$. About $30\,s$ before $C_{2}$ we expect to see three beads. Soon afterwards the bead $B3$ should vanish. By $20\,s$ before $C_{2}$ the bead $B2$ should have also almost vanished. The last inconspicuous Baily's beads will remain for another $20\,s$, fading away very slowly until the onset of totality. After third contact Baily's beads should reappear at a noticeably faster rate. After $10\,s$ there should already be three beads of moderate size. The photosphere will brighten up into a small arc by $30\,s$ after $C_{3}$. 

The computations support the broad evolution of the flash spectrum. We see thin bands of photospheric continuum shrinking and disappearing or reappearing and widening up in correspondance with these Baily's beads. The dynamical simulation especially shows how the last bead before $C_{2}$ lingered for so long and faded away so slowly.

\pagebreak

\subsection{Simulating light curves}

\begin{figure}[tb!]
\plotone{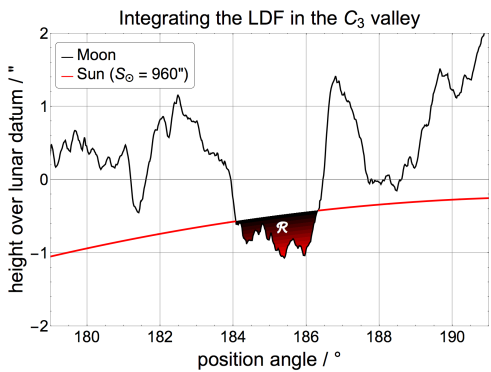}
\caption{Example integration of the limb darkening function (Equation \ref{eq:ldf}) showing the region $\mathcal{R}$ of integration where the photosphere is exposed in the $C_{3}$ valley at $T_{0} + 15\,s$.}
\label{fig:ldf_integration}
\end{figure}

\begin{figure}[tb!]
\plotone{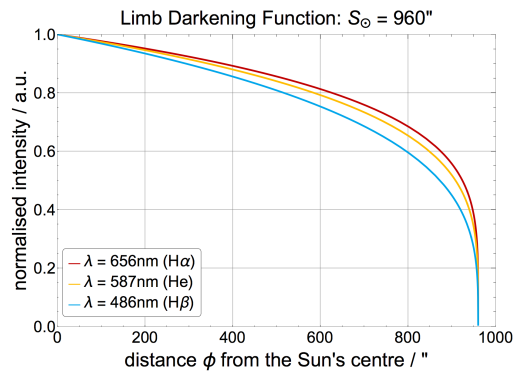}
\caption{Limb darkening function ($LDF$) at the wavelengths of the $H\text{-}\alpha$, $He$ and $H\text{-}\beta$ lines. The functional form (Equation \ref{eq:ldf}) is by Hestroffer and Magnan \citep{1998A&A...333..338H}.}
\label{fig:ldf}
\end{figure}

Light curves for the last and first Baily's beads can be extracted from the video of the flash spectrum. These observed light curves provide a quantitative measurement that we can attempt to model. 

To simulate the light curves we need to determine the area of photosphere that is still uncovered by the lunar limb. Figure \ref{fig:ldf_integration} depicts the situation. At any time we can determine the mutual position of the lunar and solar limbs and then the shape of the area $\mathcal{R}$ over which the computation is performed.

The light intensity of the photosphere is not uniform but decreases from the centre of the solar disk to its limb. The function describing the change in intensity is called the \emph{limb darkening function} (LDF). Its functional form, according to \citep{1998A&A...333..338H}, is provided by Equation \ref{eq:ldf}, where $\phi$ is the angular distance from the Sun's centre, $\Sigma_{\astrosun}$ is the (topocentric) solar semidiameter and $\lambda$ is the wavelength measured in $\mu m$. Figure \ref{fig:ldf} shows a plot of the LDF for three specific wavelengths, of the $H\text{-}\alpha$ line, the main $He$ line, and the $H\text{-}\beta$ line.

\begin{equation}
\begin{aligned}
LDF_{HM98}(\phi, \lambda) &= 
\left[
1-
\left(
\frac{\sin\,\phi}{\sin\,\Sigma_{\astrosun}}
\right)^{2} 
\right]^{\frac{1}{2}\alpha(\lambda)} \\
\alpha(\lambda) &= -0.023+\frac{0.292}{\lambda}
\end{aligned}
\label{eq:ldf}
\end{equation}

The numerical integration is set-up and performed in \emph{Wolfram Mathematica} \citep{Mathematica2018}. It uses the time polynomials, computed previously, describing the apparent topocentric lunar and solar semidiameters and the $(x,y)$ coordinates of the Sun's centre of mass (it is also the origin of the LDF). By fixing a specific time $t$ and combining those polynomials and the functional form of the LDF, we can model the light intensity originating from the uncovered area $\mathcal{R}(t)$ of photosphere by performing the integral:

\begin{equation}
I(t) = \iint_{\mathcal{R}(t)} LDF_{HM98}\, d\mathcal{R}
\label{eq:integral}
\end{equation}

Integrations are performed separately for the light curve corresponding to the $C_{2}$ valley and to the $C_{3}$ valley. The computations are done for a set of times broadly spanning the lead up to totality, totality and afterwards, at $1$-second intervals.

We would like to highlight the fact that the limb of the photosphere is considered to be sharp in these calculations. This is not physically true but all eclipse computations make that implicit assumption.

\section{Results}

\subsection{The naked eye view of the eclipse}

The progression of the eclipse at the limit of the umbral shadow path is markedly different from the progression of the eclipse near the centreline. We would thus like to give an account of the experience of totality from the limit because, besides being interesting in itself, we are only aware of a few such descriptions in the literature and it will offer some useful insights to interpret the dynamics of the flash spectrum.

The experience that most eclipse chasers (almost always stationed close to the centreline) have of the last minute before second contact is of a rather rapid and accelerating sequence of phenomena -- the level of ambient light decreases faster and faster, the lunar shadow materialises more and more noticeably and comes rushing towards the observer, the photospheric crescent shrinks and breaks into rapidly vanishing Baily's beads, often one last bright point of light shimmers past the lunar edge at the same time as the solar corona fully comes into view and then totality begins. The preceding dramatic and rapid progression of events is followed by the rather calm, albeit spectacular, phase of totality that evolves far more subtly.

From the limit this familiar experience is upended and turned into a rather alien experience. We would like to describe what we saw from observing the 2017 August 21\textsuperscript{st} total solar eclipse from just a few hundred meters inside the southern limit of the umbral shadow path. A couple of minutes before second contact, the progression of the eclipse started diverging from the common picture usually reported by seasoned observers. The ambient light level decreased organically without noticeable acceleration -- it was as if the entire sky, almost without noticing, took a deep purplish-blue tinge. Instead of an acceleration of transient phenomena, the exact opposite occurred -- a deceleration of transient phenomena. The mad rush of the breaking of the photospheric crescent into rapidly vanishing Baily's beads that happens over $10\text{-}15\,s$ at the centreline got extended to over a minute or more. From our observing site we observed for $30\text{-}35\,s$, with the naked eye and without eclipse glasses, three large Baily's beads slowly evolve, shrink and vanish on the lead up to second contact. Initially they were as bright as Venus. Then like Jupiter. Then less and less. Two Baily's beads evolved and vanished at an average rate but the last one lingered on for $10\text{-}15\,s$ after the other ones had vanished. During this gentle and subtle evolution of Baily's beads, the most striking and totally unexpected feature we observed was that the whole corona had already fully come into view. Not just the inner corona but also the outer corona extending out to several solar radii. We were busy setting up the camera so we do not know exactly when the full corona appeared but it was at least $35\text{-}40\,s$ before second contact (when we left the camera running by itself and we were just observing). The view of the slowly evolving Baily's beads on the backdrop of the full solar corona was absolutely mesmerising and enthralling. The last Baily's bead did not vanished with a bang in a diamond ring effect. It just faded away very slowly. There was no noticeable brightening of the solar corona after the disappearance of the last Baily's bead. The full corona was just there all along. After around $10\text{-}15\,s$, during which the photosphere was totally extinguished, new Baily's beads came back into view, more rapidly than the beads which had faded away before second contact. Again, no noticeable change was detectable in the brightness of the corona. After another $10\text{-}15\,s$ of unfiltered naked eye observation we averted our gaze and put on eclipse glasses for eye safety.

As shown in Figure \ref{fig:dynamic_C2C3}, the very different experience of totality from the edge is due to the peculiar way the lunar and solar limbs are travelling with respect to each other. They stay very close and they move almost tangentially to each other.

We would like to highlight something that will seem so trivial to many eclipse chasers but which will illuminate the understanding of the experience of a total eclipse from the edge of the umbral shadow path. We know that the solar corona is always there, at all times, but that the brightness of the sky is almost always so intense to hide it away from view. When the Moon completely occults the photosphere, the brightness of the sky becomes lower than the brightness of the solar corona and the corona becomes visible. We usually define the period when the photosphere is completely out of view (photospheric extinction) as totality and to approximately identify it with the period when the solar corona is fully visible. Near the centreline this is basically the case as the solar corona only fully emerges into view a couple of seconds before second contact and almost entirely disappears out of view few seconds after third contact. The photosphere is too intense at all other times to allow a clear view of the corona. We can try to artificially cover the re-emerging photosphere just after third contact and still see some traces of corona but that is dangerous and unsafe. On the contrary, very near the edges of the umbral shadow path, this is not really true anymore. As we described, by observing from just a few hundreds meters within the southern limit, despite the period of photospheric extinction being only less than about $15\,s$ long, the solar corona (including the faint outer corona) was fully visible for about $50\text{-}60\,s$. That means that the brightness of the sky had dropped below the brightness of the outer corona for far longer than the mere period of totality. We feel that we should coin a new term, \emph{coronality}, alongside \emph{totality}. Coronality is the period when the brightness of the sky falls below the brightness of the outer corona. Totality is the period when the Moon completely occults the photosphere. On the centreline (and almost anywhere else in the umbral shadow path), the two terms are almost synonyms. Very close to the edge, the two terms part ways and coronality becomes substantially longer than totality.

\subsection{Estimation of the eclipse solar radius by visual analysis of the flash spectrum video}

Our first estimate of the eclipse solar radius is based on visually inspecting the flash spectrum video to assess the duration of photospheric extinction. From that estimate we can then infer the value of the solar radius by reading it off Figure \ref{fig:duration_distance} (a). 

This is not our main methodology (see the next sub-section) but we present this simple procedure as a way of showing that Auwer's value (Equation \ref{eq:auwers_radius}) can be ruled out as being too small to be compatible with observations, even without having to perform complex numerical integrations.

\begin{figure}[tb!]
\plotone{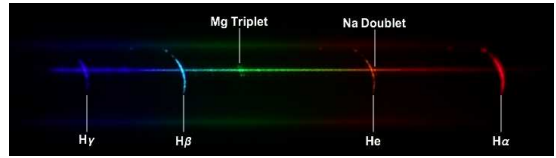}
\caption{Identification of the main features seen in a frame of the flash spectrum video recorded just south of Vale during the 2017 August 21\textsuperscript{st} total solar eclipse. Cropped from full video frame and rotated.}
\label{fig:frame_lines}
\end{figure}

Inspection of the video shows the usual features of the flash spectrum, as Figure \ref{fig:frame_lines} depicts. The resolution is not as high as what can be achieved with still images (see Figure \ref{fig:flashspectrum_details}, for example) but nevertheless the main spectral signatures present in the eclipse light can be clearly seen.

The video shows both the photospheric light of three Baily's beads fading away before second contact and the photospheric light of three Baily's beads reappearing after third contact. We expect to see this evolution based on the dynamical simulations depicted in Figure \ref{fig:dynamic_before_C2_after_C3}. In that figure the valleys generating these Baily's beads are indicated. By $20\,s$ before second contact, there is basically only light coming from the very last Baily's beads and that light fades away very slowly. The simulation confirms this slow fading away. The video then shows Baily's beads reappearing at a faster rate after third contact.

\begin{figure}[tb!]
\plotone{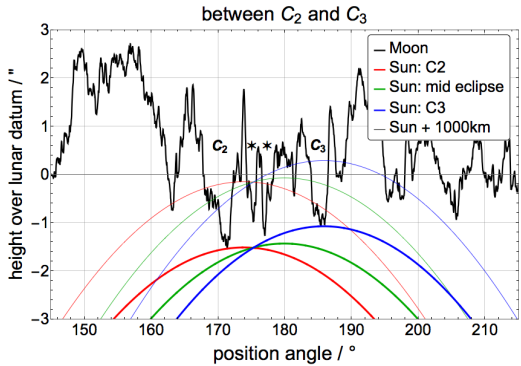}
\caption{Diagram displaying the relative positions of the lunar and solar limbs at second contact, mid-eclipse and third contact. The assumed solar radius $\sr$ is $960"$. The thinner lines indicate the locus of points located $1000\,km$ higher than the solar limb. (J. Irwin)} 
\label{fig:dynamic_moon_sun_between_C2_and_C3} 
\end{figure}

During totality there are some interesting characteristics that can be noted. The emission arcs from Hydrogen and Helium stay almost unchanged, just slightly rotating to a different position. The most interesting observation is that the light from the Magnesium Triplet never disappears either. The same is also true for the faint light from the Sodium Doublet, next to the Helium arc. The beads that can be seen during totality at the wavelength of the Magnesium Triplet indicated that the limb of the Moon was not able to travel too far past the limb of the photosphere. We know that according to Mitchell \citep{1947ApJ...105....1M} Magnesium is present from the Sun's surface up to \textapprox$3000\,km$. The bulk of the density is probably at a lower height. Let's take a height of $1000\,km$, assume a solar radius $\sr = 960"$ and simulate the dynamics of this locus of points $1000\,km$ higher than the photosphere. Figure \ref{fig:dynamic_moon_sun_between_C2_and_C3} depicts this situation. Despite the limb of the photosphere gliding under the double valley indicated by the stars, the lunar limb was never able to occult the light generated by the Magnesium Triplet. Throughout totality we can see an evolution of the intensity of these beads of light from the Magnesium Triplet. Initially the light floods the $C_{2}$ valley and a bit of the other two, then light starts flooding the $C_{3}$ valley and diminishes in intensity in the $C_{2}$ valley.

\begin{figure*}[tb!]
\gridline{
        \fig{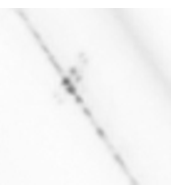}{0.2\textwidth}{(a) $t_{0} - 8.5\,s$}
        \fig{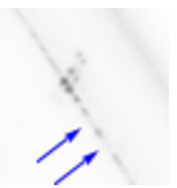}{0.2\textwidth}{(b) $t_{0} - 6.5\,s$}
        \fig{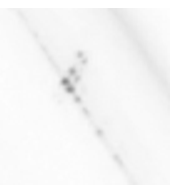}{0.2\textwidth}{(c) $t_{0} - 4.5\,s$}
         }
\gridline{
        \fig{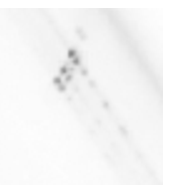}{0.2\textwidth}{(d) $t_{0} + 4.5\,s$}
        \fig{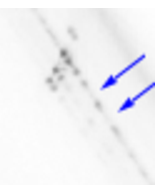}{0.2\textwidth}{(e) $t_{0} + 6.5\,s$}
        \fig{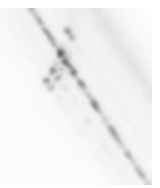}{0.2\textwidth}{(f) $t_{0} + 8.5\,s$}
         }
\caption{Sequence of zoomed-in images of the region around the Magnesium Triplet at various times. To enhance the visibility of the emission lines, the images were converted to gray scale and the colours reversed. In images (a)-(c) we can see the photospheric continuum (indicated by the blue arrows) disappearing in between emission lines (darker beads) from the fluctosphere. In images (d)-(e) we can see the photospheric continuum (indicated by the blue arrows) reappearing in between emission lines (darker beads) from the fluctosphere. $t_{0}$ is an arbitrary time corresponding to a specific reference frame in the video (Equation \ref{eq:time_transformation}).}
\label{fig:mg_triplet}
\end{figure*}

Now that we have understood some of the features seen in the flash spectrum video, we can focus on our main task -- estimating the duration of totality. We need to detect when the photospheric continuum disappears and when it reappears in order to estimate the duration of totality, that is to say, of complete photospheric extinction. Instead of looking at a whole frame, we can focus on a small region of the spectrum and inspect the video frame by frame. We select the area around the Magnesium Triplet as it is feature rich. To make the signal more contrasted, the video frames were first converted to gray-scale and then colours negated. Figure \ref{fig:mg_triplet} present a sequence of zoomed-in images of that spectral region. The video was not recorded with a UTC time signal, so we have chosen one frame situated more or less at mid-totality and referenced temporally all other frames with respect to this one using the known frame rate:

\begin{equation}
t = t_{0} + (n - n_{0}) / 23.976    
\label{eq:time_transformation}
\end{equation}

\noindent where $n$ is the frame number of a generic frame, $n_{0}$ the frame number of the reference frame and $t_{0}$ is the (unknown) UTC corresponding to the reference frame. We should note that Equation \ref{eq:time_transformation} makes the implicit assumption that no frames were dropped, which is reasonable as we used a fast SD card and we did not employ the fastest frame rate available on the camera.

The Magnesium Triplet is clearly visible towards the top of every image. In line with the $C_{2}$ valley we can see photospheric continuum in frames (a) and (f) of Figure \ref{fig:mg_triplet}. The faint darker beads that we can see to the right of the Magnesium Triplet are emission lines from the fluctosphere. If you refer back to Figure \ref{fig:flashspectrum_details}, you can identify those as emissions from Chromium and Iron. We want to look \emph{in between} those emission lines and try to see when the photospheric continuum vanishes. Blue arrows indicate what to look at. We can estimate that photospheric extinction lasted for about $13\,s$, possibly slightly less.

The photospheric continuum is present in frames (a) and (f) of Figure \ref{fig:mg_triplet} and it is absent in frames (c) and (d). The duration of totality is between these two cases, $9\,s$ and $17\,s$, with a possible value of $13\,s$. The solar radius $S_{\astrosun}$ compatible with a duration of $13\,s$ is about $960"$. This estimate has elements of uncertainty due to the presence of a background of faint emission from the fluctosphere but one thing is certain, the standard radius (Equation \ref{eq:auwers_radius}) is wildly incompatible with the duration of totality that can be extracted from the flash spectrum video. If $\sr = 959.63"$, photospheric extinction should have lasted for $32.6\,s$ at the observing site, which is clearly not the case.

Our first result is that visual inspection of the flash spectrum video supports an eclipse solar radius $\sr$ of around $960"$. By interpreting the flash spectrum video in light of the very precise eclipse computations performed here, we can provide the following estimate of the solar radius based on the range of durations given above. A duration of $9\,s$ gives $\sr = 960.09"$ and a duration of $17$ s gives $\sr = 959.93"$. Therefore, our estimate of the eclipse solar radius obtained from visual analysis can be given as:
\begin{gather}
\colorboxed{red}{\sr = (960.01\pm 0.08)"}
\label{eq:visual_radius}
\end{gather}

\pagebreak

\subsection{Estimation of the eclipse solar radius by light curve simulation} \label{sec:gamma}

\begin{figure}[tb!]
\plotone{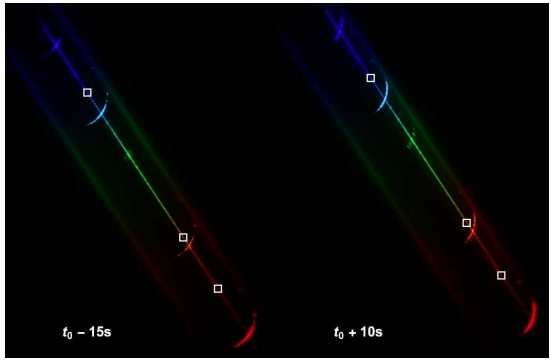}
\caption{Light curves are extracted from spectral regions (indicated with white boxes) around wavelengths $\lambda = 480\,nm$ (blue), $\lambda = 580\,nm$ (yellow) and $\lambda = 620\,nm$ (red). These regions are in line with the spectra of the light from the $C_{2}$ and $C_{3}$ valleys, where the last Baily's bead (left half of the figure) disappears at $C_{2}$ and the first Baily's bead (right half of the figure) reappears at $C_{3}$. $t_{0}$ is an arbitrary time corresponding to a specific reference frame in the video (Equation \ref{eq:time_transformation}).}
\label{fig:extraction_region}
\end{figure}

\begin{figure*}[tb!]
\gridline{
        \fig{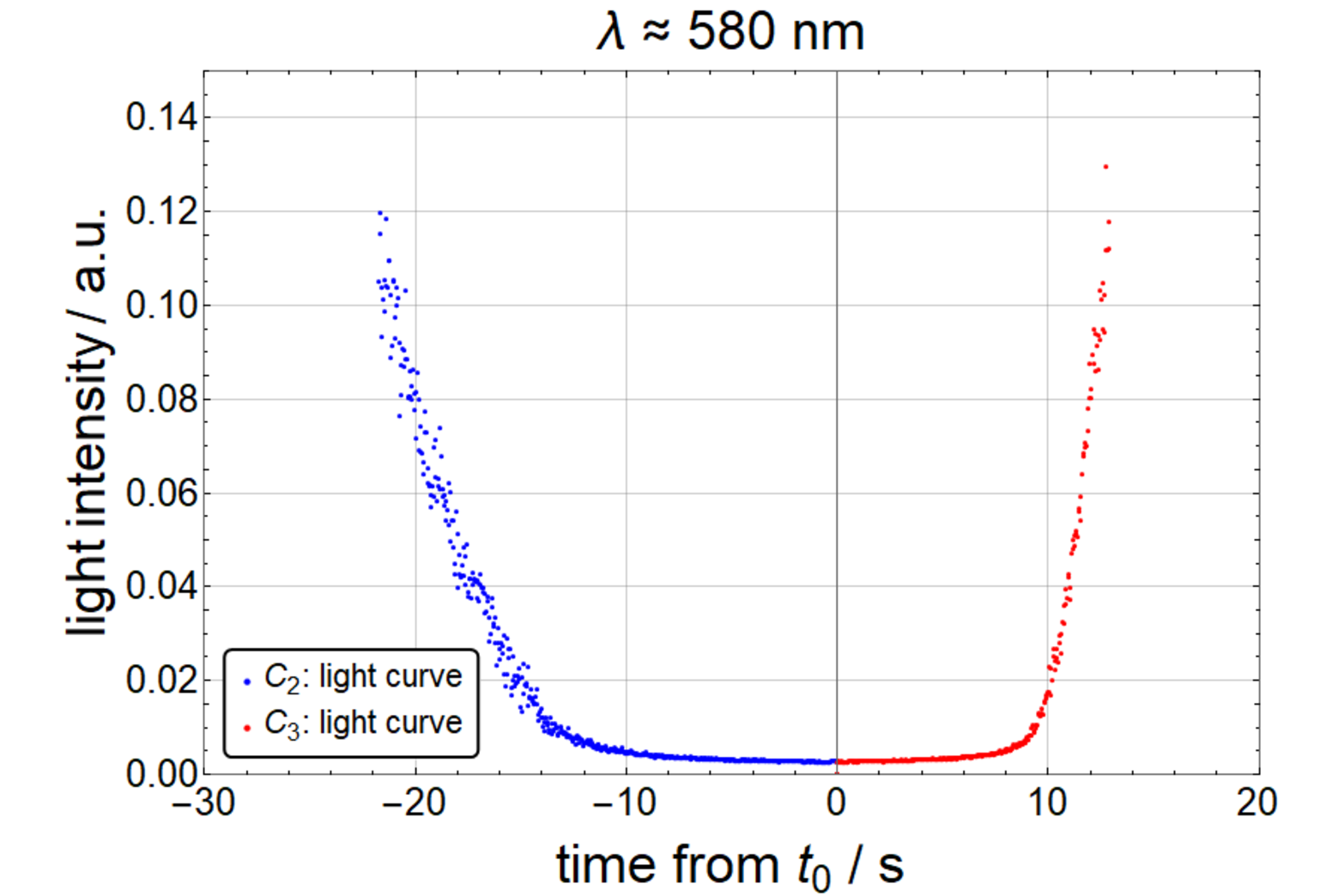}{0.45\textwidth}{Observations}
        \fig{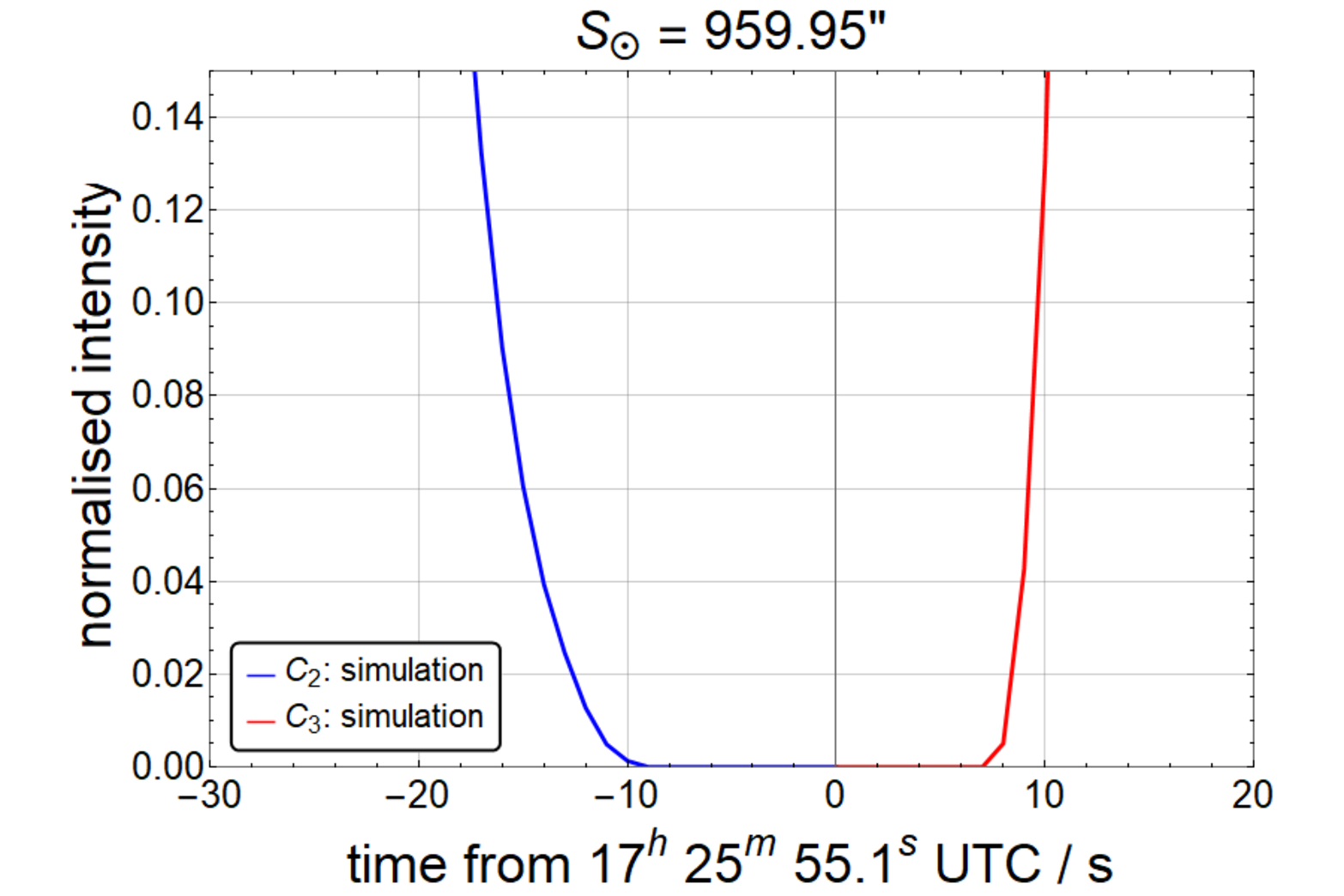}{0.45\textwidth}{Simulations}
        }
\caption{The observed light curves at $\lambda\approx 580\,nm$ and an example of simulated light curves for $\sr = 959.95"$. The time scale of the observed light curves is arbitrary and can be shifted to allow comparison. Because the LDF is normalised, a scale factor can be applied to the simulated light intensities to be able to compare them to the observed ones. Moreover, a constant bias will have to be applied to the simulated light curves to account for some background intensity in the observed light curves, extraneous from the intensity of the photospheric continuum. (L. Quaglia)}
\label{fig:raw_data}
\end{figure*}

The flash spectrum video contains sufficient data to allow us to conduct a more quantitative analysis to estimate the value of the eclipse solar radius. The idea is to extract light curves for the last Baily's bead before $C_{2}$ and for the first Baily's beads after $C_{3}$. The main spectral region chosen to perform the extraction is around $580\,nm$ and it is indicated in Figure \ref{fig:extraction_region} by the middle white boxes. The region just to the left of the Helium arc is reasonably quiet with respect to the fluctosphere, as it can be assessed by looking at the high resolution spectrum in Figure \ref{fig:flashspectrum_details}. 

The measurement of light curves from the video is based on extracting a sequence of consecutive frames and on analysing them to gather light intensity information. Frames are extracted using the \emph{OpenCV} library \citep{OpenCV2015} which imports and transforms the video data into the required $sRGB$ colour space. Given the triplet of values $\{R', G', B'\}$ for each pixel, with components expressed in the range $[0, 1]$, the linearised luminance $Y$ is obtained from the transformation:

\begin{equation}
    Y = 0.2126\,\Gamma(R') + 0.7152\,\Gamma(G') + 0.0722\,\Gamma(B')
\label{eq:luminance}
\end{equation}

\noindent where

\begin{equation}
    \Gamma(x) = 
        \begin{cases} 
            \frac{x}{12.92} & x \leq 0.04045 \\
            \left(\frac{x+0.055}{1.055}\right)^{2.4} & x > 0.04045 \\
       \end{cases}
\label{eq:gamma_decompression}
\end{equation}

\noindent The operation $\Gamma(x)$ undoes the gamma adjustment and linearises the $RGB$ components. Both equation \ref{eq:luminance} and \ref{eq:gamma_decompression} are defined in the international standard defining the $sRGB$ colour space \citep{IEC61966-2-1:1999}.

For each frame we then compute the average linearised luminance of the pixels in the extraction region (size $11\times11$ pixels). The resulting light curve can be seen in Figure \ref{fig:raw_data} (left plot). Timing is obtained by using the same relationship as in Equation \ref{eq:time_transformation}.

The intensity of the light vanishing at the bottom of the $C_{2}$ valley decreases almost linearly and then flattens out. Correspondingly, the intensity of the light reappearing at the bottom of the $C_{3}$ valley is initially almost flat and then starts increasing also almost linearly. We observe that the two light curves do not exactly reach zero intensity, even at mid-totality. We think that there might be two reasons for such behaviour. First, the flash spectrum mostly collects the eclipse light but there is some, albeit very minimal, diffuse ambient light which is probably emitted at all frequencies. So we would expect some background intensity equally affecting all light curves. Second, we know that the light from the fluctosphere is in line with the photospheric continuum. Even if we have chosen a quiet area of the flash spectrum in that regard, there are probably always traces from faint emissions from the fluctosphere. The very gentle sloping of the flat parts of both light curves might be a sign of that. Nevertheless, the bulk of the intensity in the linear sections must be due to the photospheric continuum.    

Figure \ref{fig:raw_data} shows the experimental and simulated light curves by assuming a solar radius $\sr = 959.95"$. Before being able to compare simulated and observed light curves, we need to make some adjustments. The time $t_{0}$ in the light curves is arbitrary (Equation \ref{eq:time_transformation}) so we can apply a suitable global time translation to the observed light curves to make their timescale compatible with the UTC time scale of the computations:

\begin{equation}
t_{0} \rightarrow t_{0} + \tau
\label{eq:time_shift}
\end{equation}

The LDF is normalised to $1$, so we can apply a suitable scale factor $a$ to the computed intensities to bring them on the same intensity scale the light curves are expressed in. Afterwards, we also add a global shift to the computed intensity to account for the fact that the observed light curves do not go to zero during totality (due to the background diffuse light and faint emission from the fluctosphere we pick up alongside the photospheric continuum). We apply a simple linear transformation to the simulated intensity $I$, where $a$ and $b$ are fitted parameters:  

\begin{equation}
I \rightarrow a\, I + b
\label{eq:linear_transformation}
\end{equation}

These parameters are kept constant to transform all simulations at the same wavelength, so as to allow a fair comparison. We should also note that, despite the light curves at second and third contact being independent, we were able to apply the same adjustment to both of them.  

\begin{figure*}[tb!]
\gridline{
        \fig{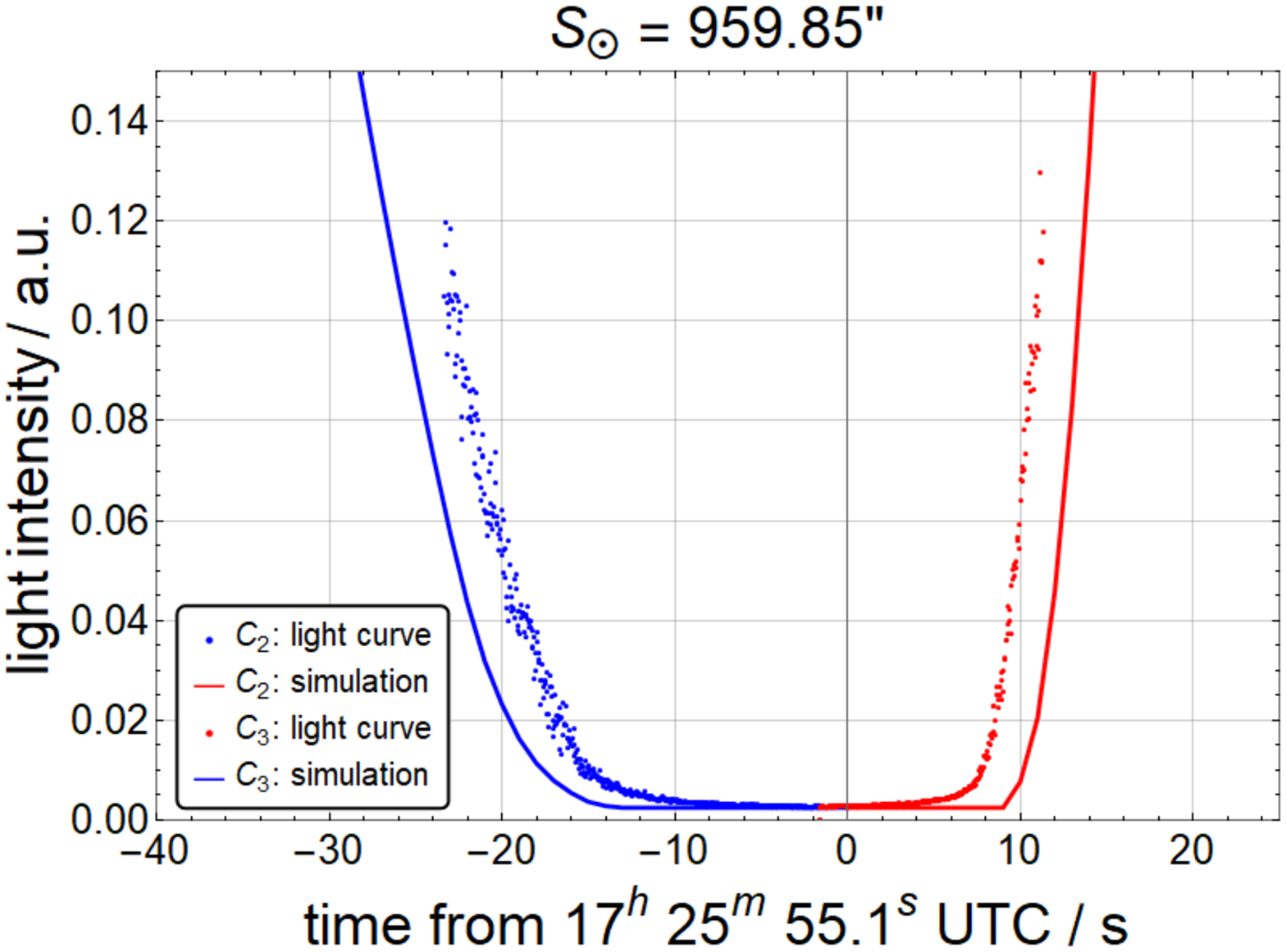}{0.45\textwidth}{(a)}
        \fig{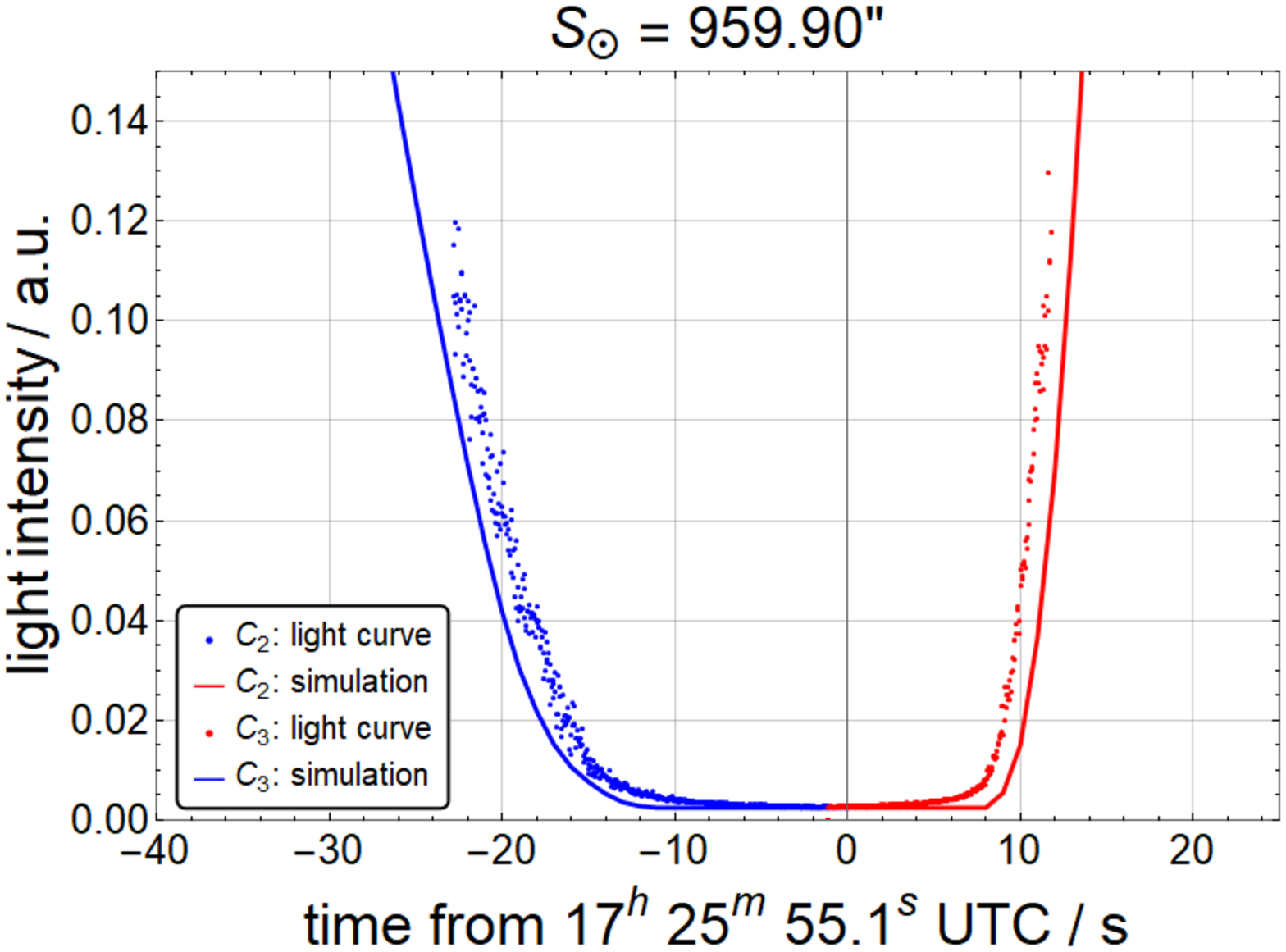}{0.45\textwidth}{(b)}
        }
\gridline{
        \fig{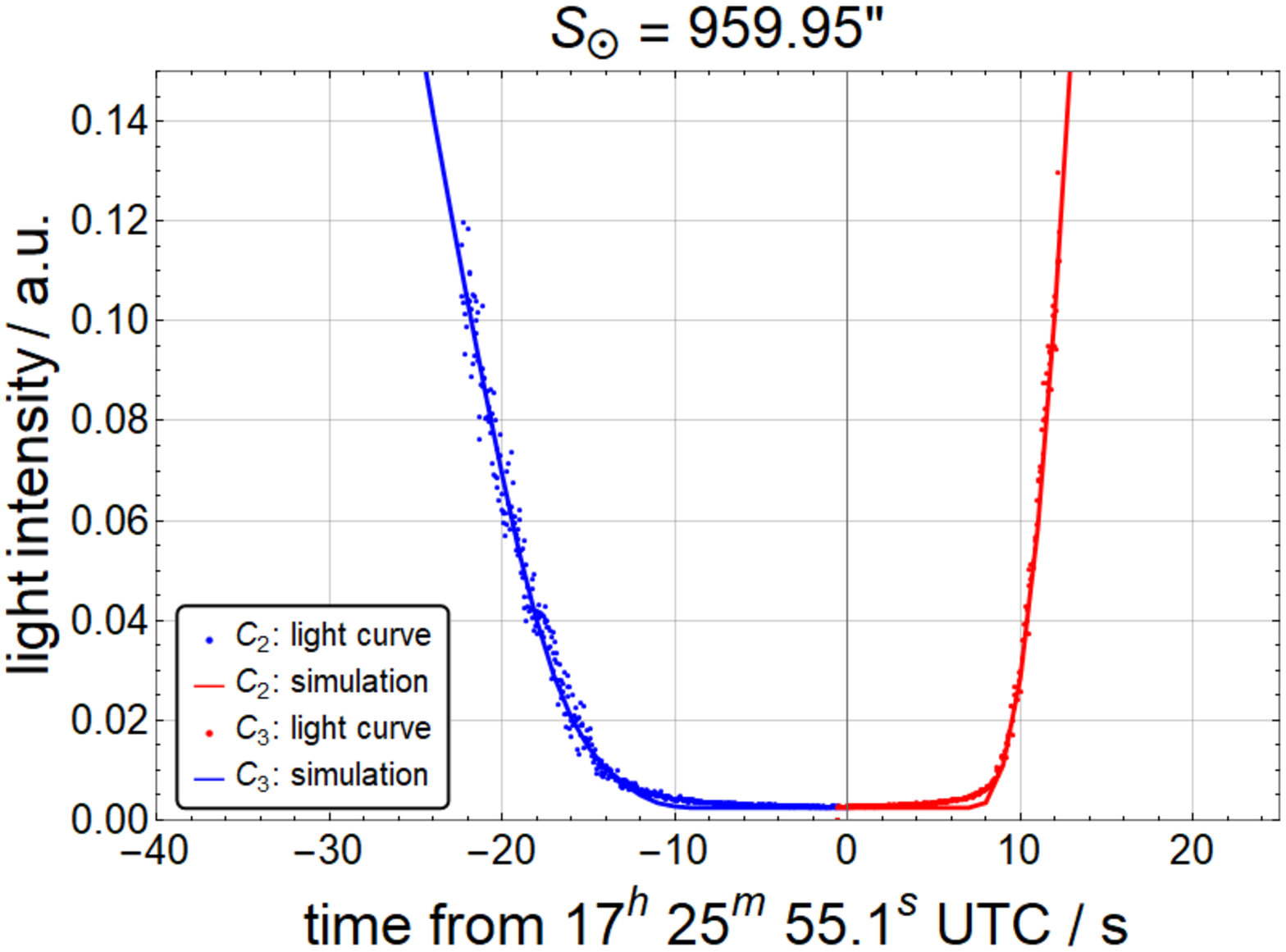}{0.45\textwidth}{(c)}
        \fig{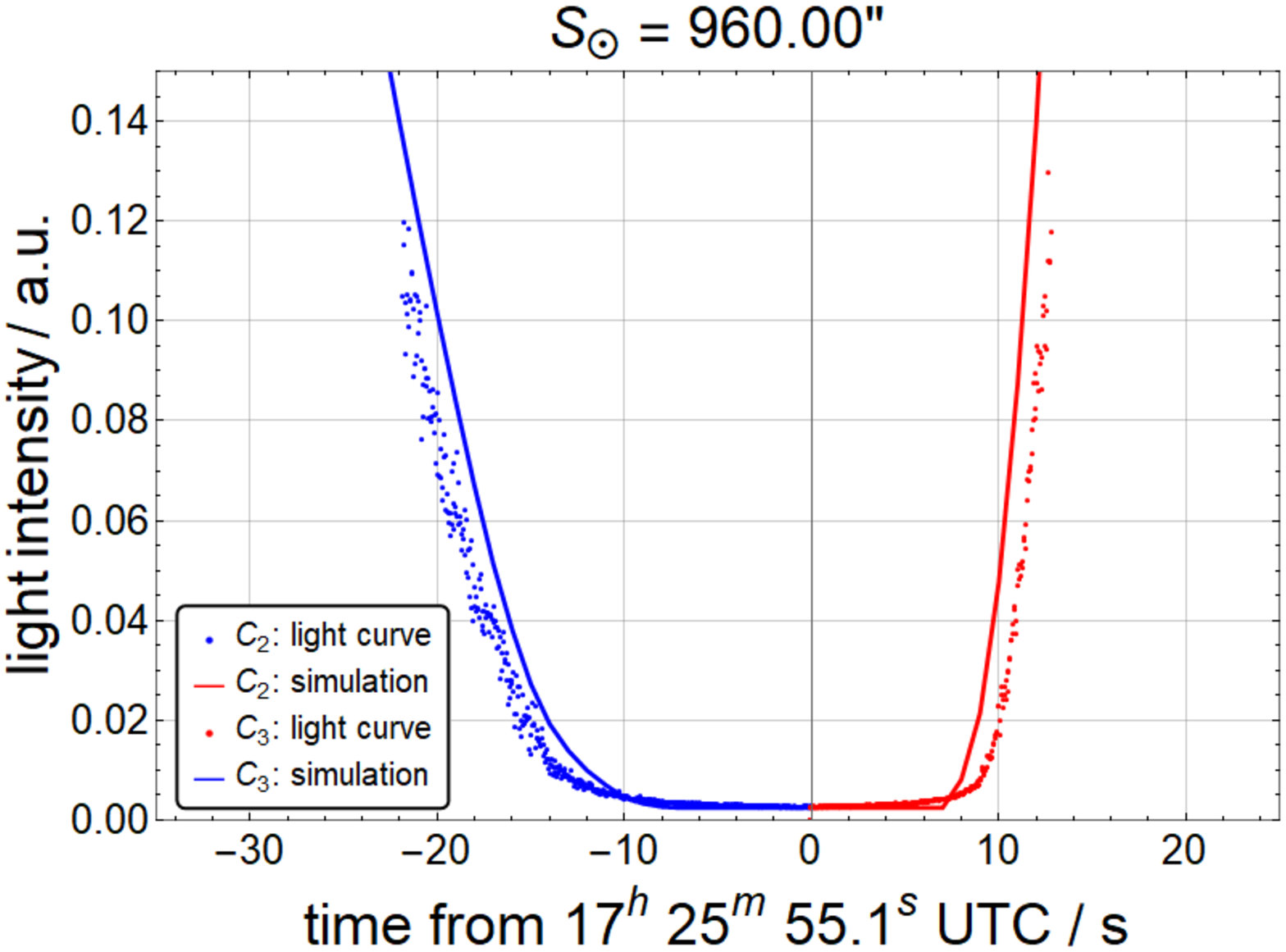}{0.45\textwidth}{(d)}
        }
\caption{Comparison of the observed light curves (at $\lambda\approx 580\,nm$) with the curves obtained from simulations using four different values of the solar radius $\sr$: (a) $959.85"$, (b) $959.90"$, (c) $959.95"$ and (d) $960.00"$. (L. Quaglia)}
\label{fig:simulations}
\end{figure*}

The results of the modelling of the light curves at $\lambda\approx 580\,nm$ are presented in Figure \ref{fig:simulations}. Four values of the solar radius $\sr$ are taken into consideration: $959.85"$, $959.90"$, $959.95"$ and $960.00"$. The best simulation of the observed light curves is obtained for $\sr = 959.95"$. The simulation is very good in the linear sections of the light curves. Small discrepancies can be seen very close to the interval of photospheric extinction. This is not surprising as the LDF ends sharply. We can take the simulations with $959.90"$ and $960.00"$ as providing uncertainty bounds for the value $\sr = 959.95"$. Therefore, our estimate of the eclipse solar radius obtained from light curve simulation can be given as:
\begin{gather}
\colorboxed{red}{\sr = (959.95\pm 0.05)"}
\label{eq:simulations_radius}
\end{gather}

\begin{figure*}[tb!]
\gridline{
        \fig{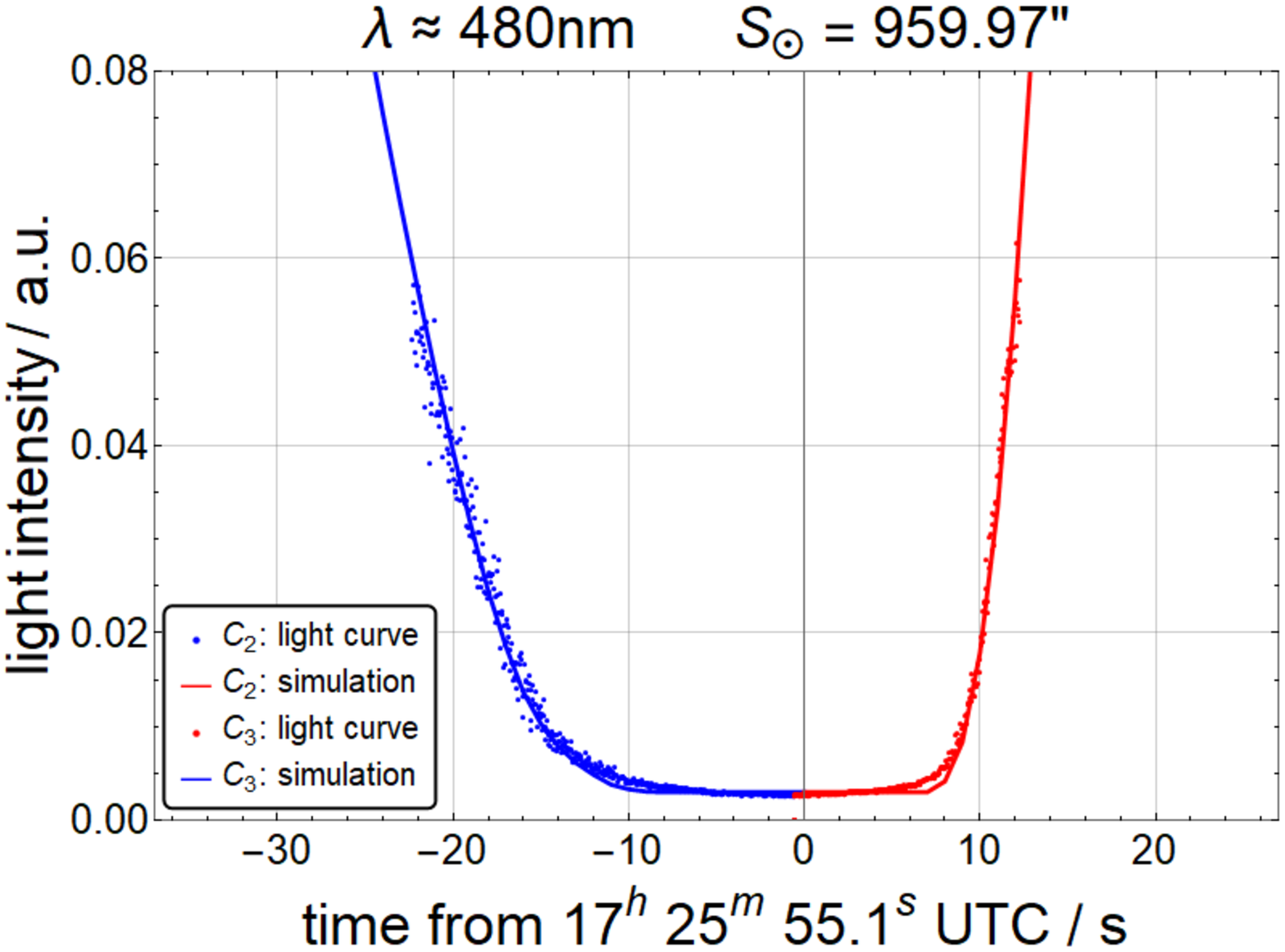}{0.45\textwidth}{(a)}
        \fig{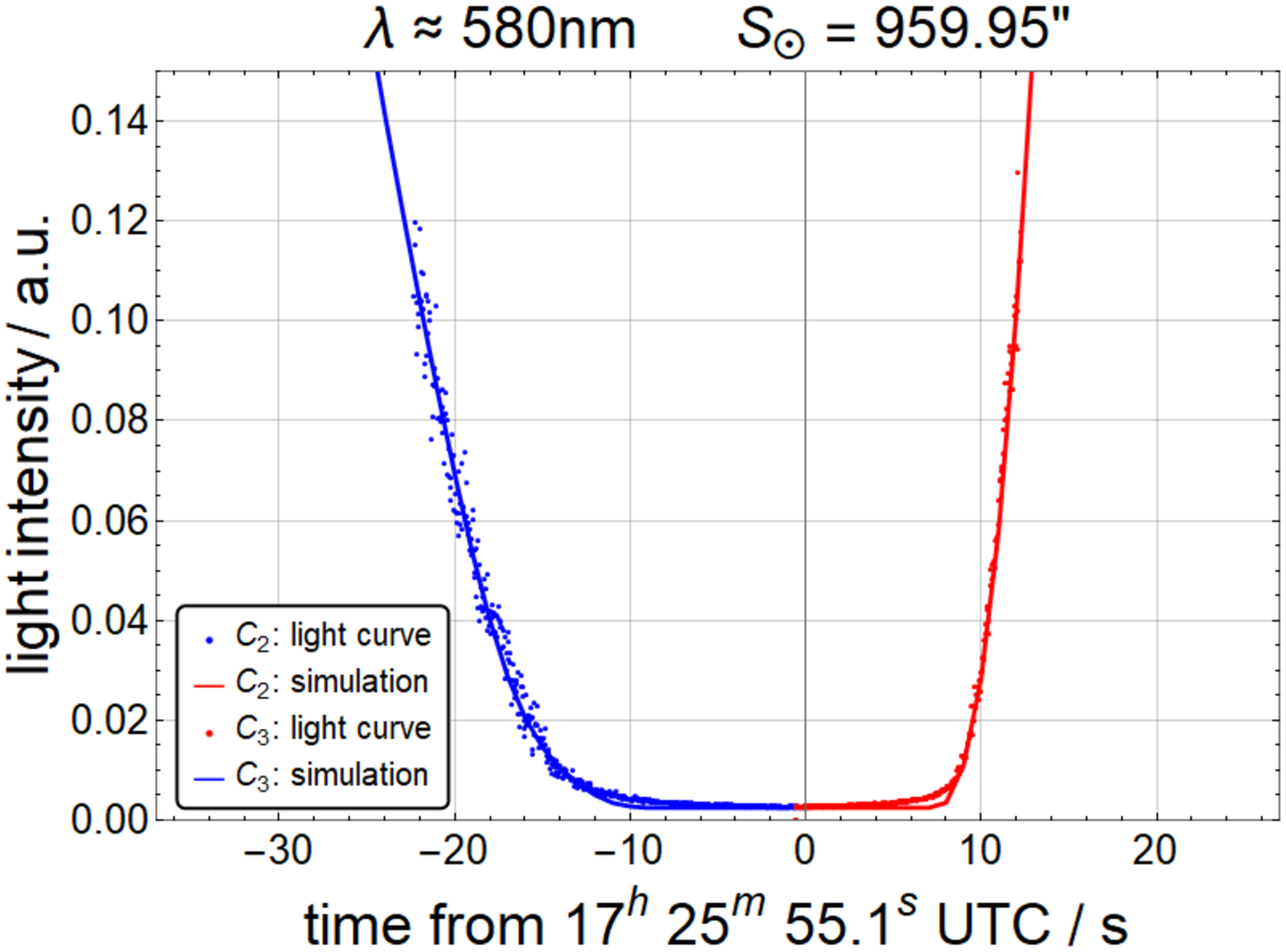}{0.45\textwidth}{(b)}
        }
\gridline{
        \fig{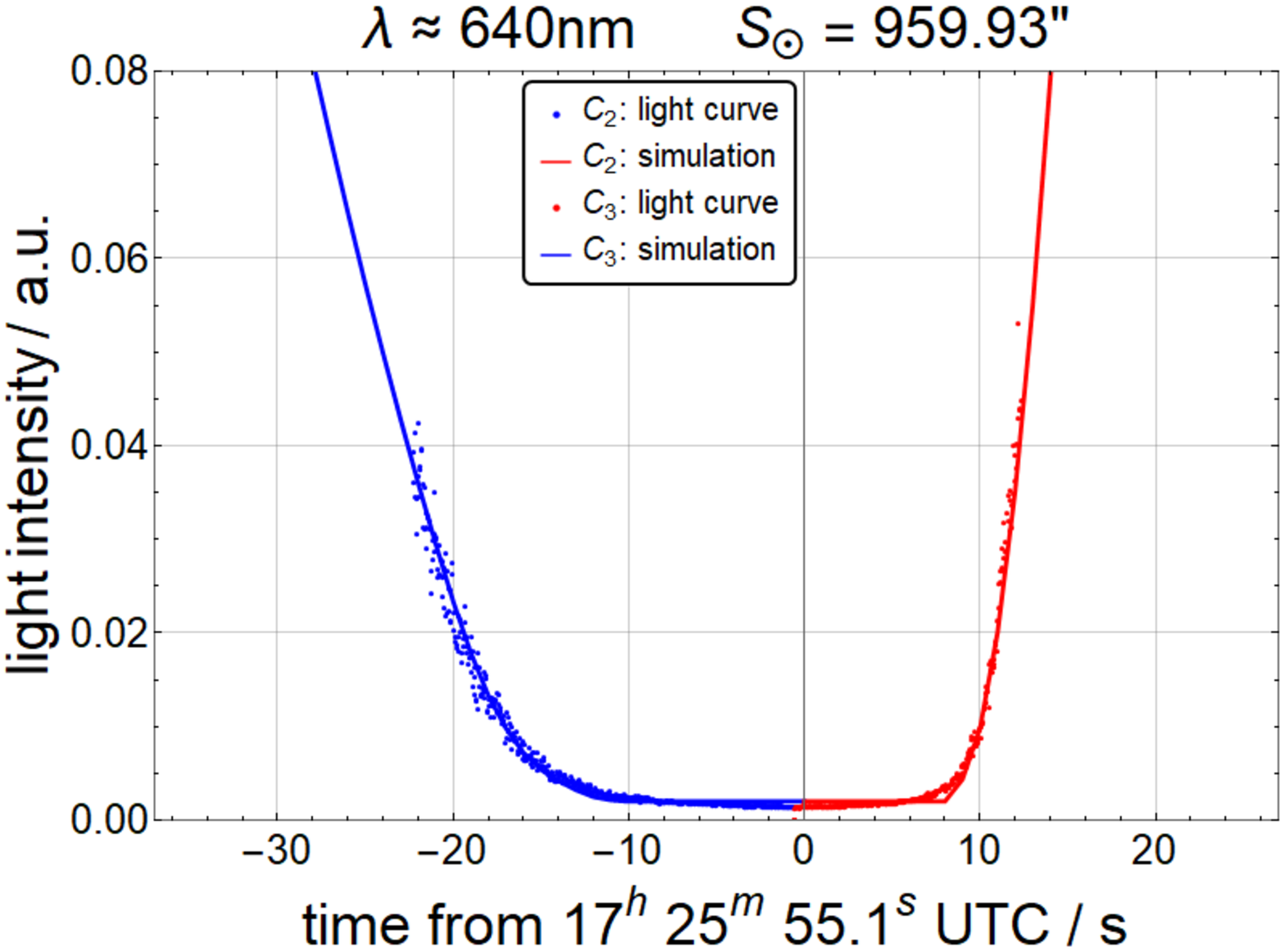}{0.45\textwidth}{(c)}
        }
\caption{Comparison of the observed light curves at three different wavelengths: (a) $\lambda\approx 480\,nm$, (b) $\lambda\approx 580\,nm$ and (c) $\lambda\approx 640\,nm$ with light curves obtained from the best fitting simulations. (L. Quaglia)}
\label{fig:simulations_wavelength}
\end{figure*}

To test if a dependency of the solar radius $\sr$ on the wavelength could be detected, we extracted light intensity curves from two other spectral regions -- one to the left of the $H\text{-}\beta$ arc at $\lambda\approx 480\,nm$, and one to the left of the $H\text{-}\alpha$ arc at $\lambda\approx 640\,nm$ (see Figure \ref{fig:extraction_region}). These light curves and their simulations are displayed in Figure \ref{fig:simulations_wavelength} together with the light curves at $\lambda\approx 580\,nm$ we have already presented. The time shift $\tau$ (Equation \ref{eq:time_shift}) is the same in all three cases. The bias $b$ (Equation \ref{eq:linear_transformation}) is very close in all three examples. The scale factor $a$ (Equation \ref{eq:linear_transformation}) had to be adjusted at different wavelengths because the camera sensor is less sensitive in the blue and in the red. In Figure \ref{fig:simulations_wavelength} we can see that a solar radius of about $\sr = 959.95"$ better fits the observations at these wavelengths. However, we do not consider the minor shift in the best fit solar radius at different wavelengths to be significant relative to the uncertainty on each individual estimated radius.

\section{Discussion}

We can assess the impact of using a solar radius different from the standard one by considering the example of the 2017 August 21\textsuperscript{st} total solar eclipse. If we take an observer on the centreline, witnessing mid-totality at the same time as our observing site, totality lasts $2^{m}\,10.0^{s}$ if we use Auwers' radius (Equation \ref{eq:auwers_radius}) and $2^{m}\,08.4^{s}$ if we use the radius estimated in this work (Equation \ref{eq:simulations_radius}) -- a change of $1.6\,s$. This change might seem small but it has practical implications. For example, some eclipse observers rely on automation to trigger cameras and change exposures as the eclipse progresses, according to preset scripts based on automatically computed contact times \citep{Maestro2019, Orchestrator2019}. One example of an application where automation is often used is the MMV (Mathematical Methods of Visualization of the solar corona) Project \citep{2006CoSka..36..131D, MMV2021}. Having access to a more accurate value of the eclipse solar radius allows computing contact times more accurately and hence will improve the performance of any automation.

\begin{figure*}[tb!]
\plotone{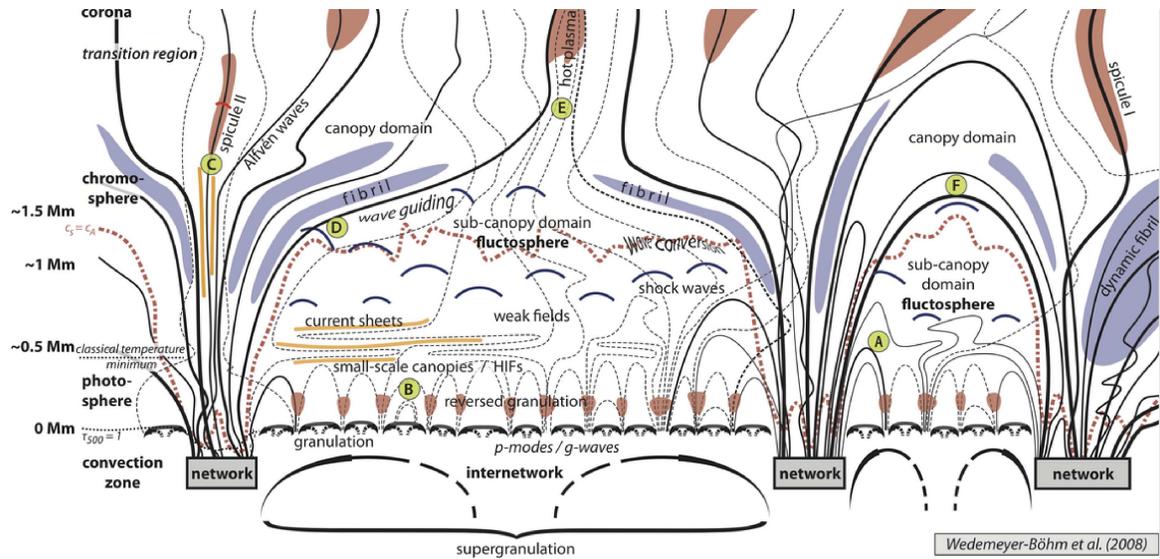}
\caption{Diagram of the transition zone, which is very complex, between photosphere and chromosphere. The photosphere morphs into the fluctosphere and then into the chromosphere proper. In a simplistic manner, we can infer that the edge of the photosphere is possibly $50\,km$ wide (source of the image: \citep{2009SSRv..144..317W}).} 
\label{fig:solar_structure} 
\end{figure*}

In all previous diagrams and computations, we have assumed a sharp solar edge. This is the assumption that all eclipse computations implicitly make. It would be impractical to do otherwise. Figure \ref{fig:solar_structure} depicts the very complex situation occurring at the interface between photosphere, fluctosphere and chromosphere. The photosphere is a layer \textapprox$500\,km$ deep and the eclipse radius should approximately be the one corresponding to the top of the photosphere. As the Sun is not a solid body, the edge of the photosphere should be thought more as a transition layer than as a well defined surface.  

Let's take the width of this transition layer to be $50\,km$ wide, as an order of magnitude, and let's assess what impact that assumption carries. If we take the solar radius to be $959.95"$ (at $1\,au$), the transition zone turns out to be \textapprox$0.07"$ wide. This inform us on the probable accuracy with which we can quote the value of the solar radius. Aiming at quoting the eclipse solar radius to $\pm 0.05"$ seems reasonable and it is remarkable that our simple measurement is able to say something close to that level of accuracy. The Sun's radius is \textapprox$7\times10^{5}\,km$ and Earth is on average \textapprox$1.5\times10^{8}\,km$ away from the Sun. Then $50\,km$ is really not very much in comparison. Nevertheless, observing total solar eclipses from the limit of the umbral shadow path provides quite a sensitive measuring procedure to estimate the solar radius.

Another consequence of the transitional nature of the edge of the photosphere is that second and third contact should be interpreted as short intervals of time rather than as single individual moments. Assuming that the edge of the photosphere is $0.05"$ thick and that it is centred at $\sr = 959.95"$, the duration of totality becomes a range and not a single value anymore: $[2^{m}\,08.3^{s},\, 2^{m}\,08.5^{s}]$. Similarly, the umbral shadow path limits should be viewed as fuzzy bands \textapprox$100\,m$ wide, rather than sharp lines.

\section{Conclusion}

This is our proposal for the eclipse solar radius as derived from the analysis of the flash spectrum of the 2017 August 21\textsuperscript{st} total solar eclipse:
\begin{gather}
\colorboxed{red}{\sr = 959.95"}
\label{eq:proposed_radius}
\end{gather}

This value can be directly used in eclipse computations, recalling that we always make the implicit assumption of having a sharp solar edge. The eclipse solar radius was found not to be dependent on wavelength.

We have presented a methodology to estimate the eclipse solar radius that brings together aspects of previous techniques: using the flash spectrum and extracting light curves \citep{1993PASJ...45..819K}; observing right at the umbral shadow path limits \citep{1994SoPh..152...97F}; and simulating light curves \citep{2015SoPh..290.2617L}.

The methodology is, within reason, easily applicable, at least where it concerns the data collection phase. Its strength relies on:

\begin{itemize}
\item observing from the umbral shadow path limits, to increase the sensitivity of the measurement on the solar radius.
\item recording the flash spectrum evolution, to disentangle the light sources present in the eclipse light and to provide access to wavelength dependent data. 
\item very accurate eclipse computations, to minimise uncertainties by precisely accounting for the complexity of the lunar limb topography and of the fine details of the eclipse geometry.
\item light curve simulations, as matching a whole curve provides tighter requirements than just fitting contact times. 
\end{itemize}

The methodology could be improved by acquiring higher resolution flash spectra to allow a greater geometric separation between the photospheric continua of nearby Baily's beads. This would provide more light curves to simulate and it would make the light curve matching more constraining.  

\section{Future work}

We will be aiming to record a higher-resolution flash spectrum video near the limits of the umbral shadow path during the 2023 April 20\textsuperscript{th} annular-total eclipse, visible from Exmouth Peninsula in Western Australia. Due to the intrinsically short nature of that eclipse, stemming from having a magnitude $A$ very close to $1$, the emission arcs in the flash spectrum will be unusually long. This should enhance all the features normally seen in the flash spectrum.

We will also aim to use larger optics and a higher quality diffraction grating to obtain higher-resolution spectra. This should provide better separation, without contamination, of the photospheric continuum generated by individual Baily's beads. This will increase the number of measurements that can be extracted from a single flash spectrum video. We will also consider the possibility of setting up multiple measuring stations, closely-spaced and straddling the theoretical umbral shadow path limits, to probe different evolutions of the flash spectrum and to increase the number of measurements. 

Moreover, we will use an imaging device that allows for fast acquisition of RAW frames, making full use of the resolution and linearity of the sensor without intervening software transformations (like the gamma adjustment). We will record a UTC time-stamped video. There are now on the market CCD cameras with very accurate in-built UTC time-stamping of every frame. Alternatively, we can use an accurate ($<1\,ms$) UTC timing device that we have developed and tested and that will work anywhere in the world. Time-stamping will make matching of the simulated and observed light curves more stringent -- besides having to reproduce the temporal spacing between the light curves, the absolute timing of each light curve will also need to be reproduced. This might then allow testing of the accuracy of eclipse computational models alongside estimating the eclipse solar radius (see Appendix C).\newline

\noindent We sincerely thank the anonymous reviewer for comments and suggestions that helped improve and clarify this manuscript. We also thank Professor Shadia Habbal (Institute for Astronomy, University of Hawai'i) for her encouragement throughout the writing of this report.

\appendix

\section{Time-stamping of the flash spectrum video}

The methodology we have presented to estimate the eclipse solar radius relies on two components:

\begin{itemize}
\item{a video of the eclipse flash spectrum recorded from very close to the limit of the umbral shadow path} 
\item{very precise eclipse computations}    
\end{itemize}

Time-stamping of the video is not strictly necessary. It would be beneficial to have it, to assess if any frames were dropped, but the matching of simulated and observed light curves does not strictly depend on it. This is one of the advantages of observing from very close to the path limits.

Even if we do not have the time-stamping, we can recover it with reasonable accuracy, \emph{post facto}. We highlight that a secondary outcome of the procedure of matching observed and simulated light curves is to obtain an estimate of the offset $\tau$ (Equation \ref{eq:time_shift}) between the arbitrary reference time $t_{0}$ (Equation \ref{eq:time_transformation}) of the observed light curves and the reference time $T_{0}$ (Equation \ref{eq:reference_time}) used in the polynomials describing the apparent topocentric semidiameters of the Sun and Moon and the relative positions of their centres of mass -- key ingredients to compute the simulated light curves. This offset $\tau$ is one of the adjustments described in Figure \ref{fig:raw_data}.

The flash spectrum video has $1440$ frames and the arbitrary reference time $t_{0}$ is attached to frame $881$. The light curve matching procedure returns an estimate of the value of $\tau = -0.55\,s\pm 0.15\,s$, so we can assign frame $881$ with the time $T_{0}-0.55\,s$, that is to say $17^{h}\, 25^{m}\, 54.55^{s}$ UTC. Based on this attribution, we can then time-stamp in UTC all other frames in the video using the known frame rate. The video thus starts at approximately $17^{h}\, 25^{m}\, 17.85^{s}$ UTC and ends at approximately $17^{h}\, 26^{m}\, 17.86^{s}$ UTC.

\section{Comparison between flash spectrum evolution and dynamical simulations}

Once the flash spectrum video is time-stamped in UTC, we can compare every frame with the corresponding simulation of the positions of the solar and lunar limbs. We created two videos showing a side-by-side comparison, one video\footnote{\url{https://youtu.be/JI2AIRI9Prk}} assuming a solar radius $\sr = 959.95"$ and another\footnote{\url{https://youtu.be/LTxSM4Goo5A}} assuming $\sr = 959.63"$.

\begin{figure*}[tb!]
\plotone{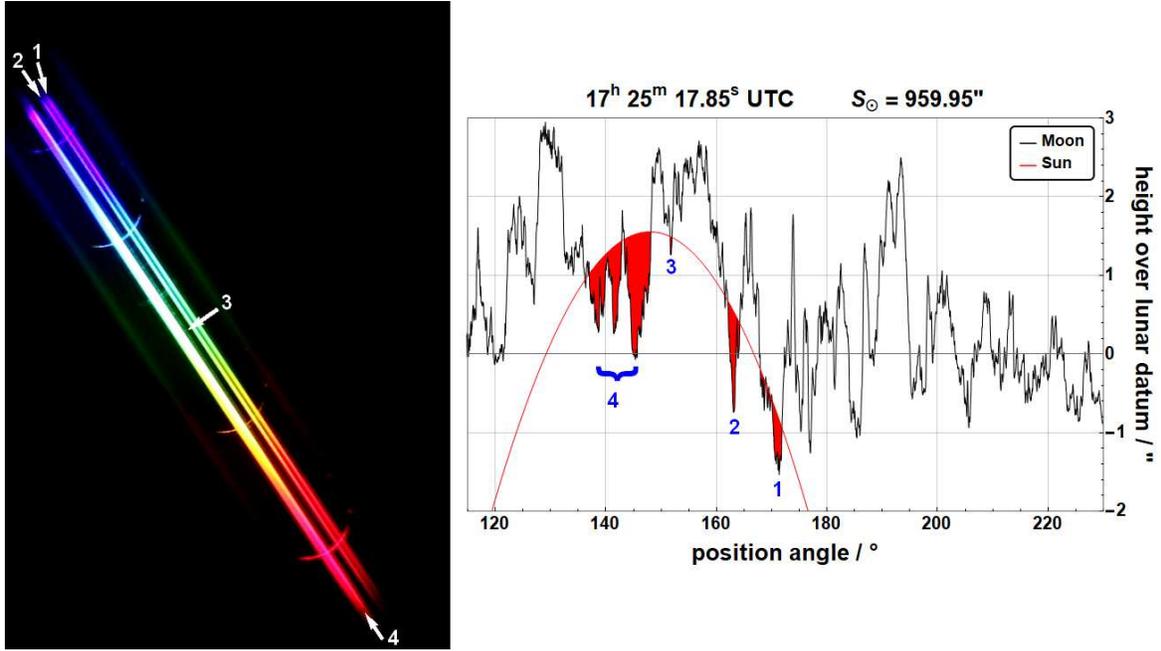}
\caption{Example composite image made possible by the \emph{post facto} time-stamping of the flash spectrum video. On the left, a frame of video, and on the right the corresponding dynamical simulation of the position of the lunar and solar limbs. We have indicated, with corresponding numbers, valleys (or group of valleys) on the lunar limb and bands of photospheric continuum. It is remarkable that the simulation reproduces the thin band of photospheric continuum originating by the narrow lunar valley indicated by the number 3. (A. Pessi)}
\label{fig:composite}
\end{figure*}

An example of side-by-side comparison is shown in Figure \ref{fig:composite} where the solar radius $\sr$ is assumed to be $959.95"$ (the value estimated in this work). In the frame we can see several bands of photospheric light -- two thin bands marked with numbers $1$ and $2$, a razor thin one marked with number $3$ and finally a thicker one marked with number $4$. The source of this light can be assessed in the simulation of the relative positions of the solar and lunar limb and corresponding numbers identify the lunar valleys generating the Baily's beads. The light coming from the valleys collectively indicated by the number $4$ coalesces together in one thick band, while the narrower valleys indicated by the numbers $1$ and $2$ generate close but separated thinner bands. The most intriguing detail that the simulation captures is the reproduction of a very thin band generated by the narrow valley indicated by the number $3$. This is testament to both the reliability of the \emph{post facto} time-stamping of the video and, even more, to the reliability of the computations.

The composite video generated by assuming $\sr = 959.95"$, our proposed eclipse solar radius (Equation \ref{eq:proposed_radius}), should be compared with the one generated by assuming $\sr = 959.63"$, the standard Auwers' radius (Equation \ref{eq:auwers_radius}). The former video accurately reproduces the evolution of the flash spectrum over the entire duration of the video. The timing of disappearance and reappearance of $8$ Baily's beads is simulated faithfully. The latter video is noticeably too early in reproducing the disappearance of Baily's beads and too late in reproducing the reappearance of Baily's beads. This is not surprising because the value of the standard radius has been found to be far too small to be compatible with eclipse observations.    

\section{Comparison of eclipse computational models}

\begin{deluxetable*}{lccc}
\tablecaption{Predicted contact times and duration at the observing site}
\label{tbl:contact_times}
\tablewidth{0pt}
\tablehead{
\colhead{Source}          & 
\colhead{Second Contact}  & 
\colhead{Third Contact}   &
\colhead{Duration}
}
\startdata
Solar Eclipse Maestro & 
$17^{h}\, 25^{m}\, 31.6^{s}$ & 
$17^{h}\, 26^{m}\, 07.7^{s}$ & 
$36.1\,s$ \\
Occult\textsuperscript{a} & 
$17^{h}\, 25^{m}\, 33.6^{s}$ & 
$17^{h}\, 26^{m}\, 06.9^{s}$ & 
$33.3\,s$ \\
Occult\textsuperscript{b} & 
$17^{h}\, 25^{m}\, 32.9^{s}$ & 
$17^{h}\, 26^{m}\, 07.0^{s}$ & 
$34.1\,s$ \\
this study (J. Irwin) & 
$17^{h}\, 25^{m}\, 34.3^{s}$ & 
$17^{h}\, 26^{m}\, 06.9^{s}$ & 
$32.6\,s$ \\
\enddata
\tablecomments{All data computed for $\sr = 959.63"$. Times in UTC.\\ (a) data from the main eclipse page (b) data from the Baily's beads analysis tool.}
\end{deluxetable*}
\vspace{-0.85cm}

We compare the predictions of our eclipse computational model to predictions provided by two publicly available applications that use the traditional method. One is \emph{Occult} \citep{Occult2021} and the other is \emph{Solar Eclipse Maestro} \citep{Maestro2019}. These applications are considered to be precise and allow the value $\sr$ of the solar radius to be varied by the user.

In Table \ref{tbl:contact_times} we present the contact times (an absolute quantity) and the duration of totality (a differential quantity) at the observing site predicted by all three models for $\sr = 959.63"$. Occult has a main eclipse page where detailed limb-corrected predictions can be obtained but also a Baily's beads analysis tool that can be used to derive contact times, so two predictions are quoted. The variability of contact times highlights the importance of developing a complete computational model where all aspects of the calculation can be made as precise as possible and where everything can be inspected.  

\begin{figure}[tb!]
\plotone{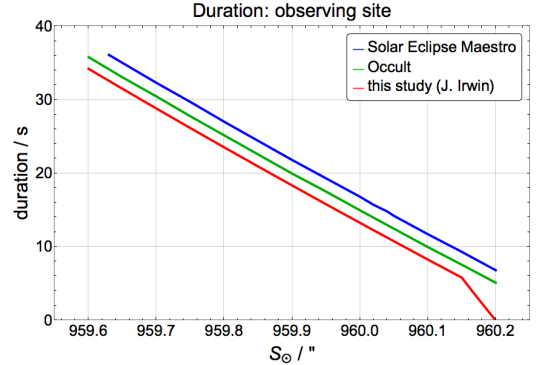}
\caption{Comparison between the duration of totality forecast for a range of values of the solar radius $\sr$ by the eclipse computational model used in this study and by two publicly available software applications for eclipse computations.}  
\label{fig:compare_durations}
\end{figure}

In Figure \ref{fig:compare_durations} we compare the duration of totality at the observing site predicted by our computational model with Solar Eclipse Maestro and Occult's Baily's beads analysis tool. For the same value of the solar radius $\sr$, Occult gives durations that are \textapprox$2\,s$ or more higher that those predicted by our model while Solar Eclipse Maestro is \textapprox$4\,s$ or more higher. Alternatively, for a given value of the duration of totality, Occult would infer a value for the solar radius $\sr$ \textapprox$0.03"$ higher than the one inferred by our model while Solar Eclipse Maestro would be \textapprox$0.07"$ higher. More interestingly, the bend in the plot at $960.15"$ (due to a change in the position of third contact on the lunar limb) is not predicted by either application. This is a hint that near the limits of the umbral shadow path the computations become very sensitive to a host of fine details for which our direct approach works seamlessly and consistently.

The final judge on the accuracy of eclipse computational models is \emph{observation}. So far it has been very hard to measure accurate contact times with sufficient precision to assess the reliability of eclipse predictions. The light curves matching technique could provide an avenue to test the models if the flash spectrum video is recorded with an accurate UTC time-stamp. This will eliminate the need (and the liberty) to perform the time adjustment described in Equation \ref{eq:time_shift}. The time-base for the observed light curves will already be UTC so the light curve simulations will have to match not only the spacing (related to the duration of totality) between the $C_{2}$ and $C_{3}$ light curves but also their absolute timing (related to the individual contact times). It should be possible to detect any discrepancy in the timings of $1\,s$ or more. This will be an additional goal for any future application of the methodology described in this paper.

\bibliography{EclipseSolarRadius_ApJS_2021}{}
\bibliographystyle{aasjournal}

\end{document}